# Stess-testing the system

## Financial shock contagion in the realm of uncertainty

**Stefano Gurciullo**

*Upgrade report submitted in partial fulfillment*
*of the requirements for the Doctoral Degree*

***University College London***

November 2014

// Abstract

This work proposes an augmented variant of DebtRank with uncertainty intervals as a method to investigate and assess systemic risk in financial networks, in a context of incomplete data. The algorithm is tested against a default contagion algorithm on three ensembles of networks with increasing density, estimated from real-world banking data related to the largest 227 EU15 financial institutions indexed in a stock market. Results suggest that DebtRank is capable of capturing increasing rates of systemic risk in a more sensitive and continuous way, thereby acting as an early-warning signal. The paper proposes three policy instruments based on this approach: the monitoring of systemic risk over time by applying the augmented DebtRank on time snapshots of interbank networks, a stress-testing framework able to test the systemic importance of financial institutions on different shock scenarios, and the evaluation of distribution of systemic losses in currency value.

# Contents

## List of Figures

## List of Tables

## List of Equations

# 1. Introduction

The recent economic and financial crisis has – more than any previous crises in the last fifty years – caused uproar against and a rejection of the conventional methods of analysis of economic phenomena, especially with regards to finance. The attacks have not come only from heterodox economics. In an interview, investor George Soros (2013) calls for a radical remaking of the discipline: this has now become too focused on a Newtonian-physics view of reality, with a mathematical formalism that is too detached from real dynamics and thus from policy. Analogously, in his books Nassim Taleb (2005, 2010, 2012) provides a detailed critique of conventional economics with specific regards to finance: models which assume general equilibrium, perfect or near-to-perfect information and steady-state systems appear to be empirically invalid, and policy actions derived from them engender a great risk of unintended consequences. The underestimation of the relevance of cascade effects in financial markets, for instance, has prevented policymakers to act preemptively in 2008, thus contributing to the economic collapse that followed (Haldane and May, 2011). Other powerful criticisms of the current conceptual and methodological stance of economics are given by Shiller (2010), Krugman (2011), Stiglitz (2012), and Cristian, Dumitru-Alexandru and Daniel (2012), to name a few.

This paper attempts to present a network-based approach to the modeling of financial crises and shocks, with great attention given to the issue of uncertainty and missing data. More specifically, it advances the augmentation of the DebtRank algorithm with uncertainty intervals, derived by the application of the fitness model to the reconstruction of financial networks. It is also demonstrated how this variant of DebtRank is superior to the standard default contagion algorithm through an experiment involving synthetic networks estimated from real data.

The paper is structured into seven main chapters, introduction included. Chapter 2 introduces the research question – namely, *how are financial systems affected by adverse events?* – and provides the motivational and literature backgrounds to the research. After a brief account of the effects of financial crises onto the wider economy and society, relevant literature will be reviewed. Three strands will be critically discussed: indicator-based approaches to systemic risk, Dynamic Stochastic General Equilibrium (DSGE) models, and network analysis, to which the research contributes. Chapter 3 presents the design of the model. It will explain how interbank systems are operationalised in the network paradigm, and will show the mathematical framework behind the main two modules of the model, respectively dealing with network reconstruction and financial shock contagion simulation.

Subsequently, Chapter 4 implements the model in an experiment setting. The experiment will test the performance of the Debtrank and default contagion algorithms on three synthetic interbank network ensembles, demonstrating that the former is able to detect an increase in the systemic importance of a bank in a more sensitive way than the latter. The results of the experiment do not only contribute to the advancement of knowledge on systemic financial risk, but also have deep policy implications, which are considered in Chapter 5. This section will briefly present three possible applications of the framework. The first is aimed at the monitoring of systemic risk over time, the second at the stress-testing of a financial institution at different conditions of distress, and the third at the evaluation of loss probability distributions referring to one or more financial institutions, conditional on distributions of shocks affecting one or more shocked banks. The latter instrument reveals to be particularly useful for the calculation of Values at Risk (VaR) estimates. Chapter 6 will briefly assess the limitations of the work. A timetable will show how the author plans to perform the subsequent research steps. Finally, concluding remarks will follow in Chapter 7.

## 2. Background

Due to the last crisis, 371 commercial banks failed between 1/1/2008 and 1/7/2011 in the United States alone (FDIC, 2013). Yet, little is still known as to why the impact has been so great. Attention to the systemic level of financial institutions has increased only following these events (Catte et al., 2010). Macroprudential policy – policy aimed at ensuring financial stability – has become a major topic of interest among policy makers and academics only then (Clement, 2010). The first part of this section presents and explicates the relevance of the research question of this work, while the second part reviews three main approaches that are tackling it.

### 2.1 The research question and its importance

The goal of this work is to explore and synthesise measures that are able to quantitatively assess the systemic importance of financial institutions in adverse scenarios. The research question, thus, is: *how are financial systems affected by adverse events?* Good answers to the question have the potential to contribute to the development of preemptive measures against financial crises. It follows that the main reason to pursue this work is the opportunity to limit the negative consequences of financial distresses onto the wider economy and society.

Economic history seems to agree to the statement that financial crises are detrimental to society. Reviews by Reinhart and Rogoff (2008; 2009) show stylised facts about the effect of financial turmoil on GDP growth and unemployment.

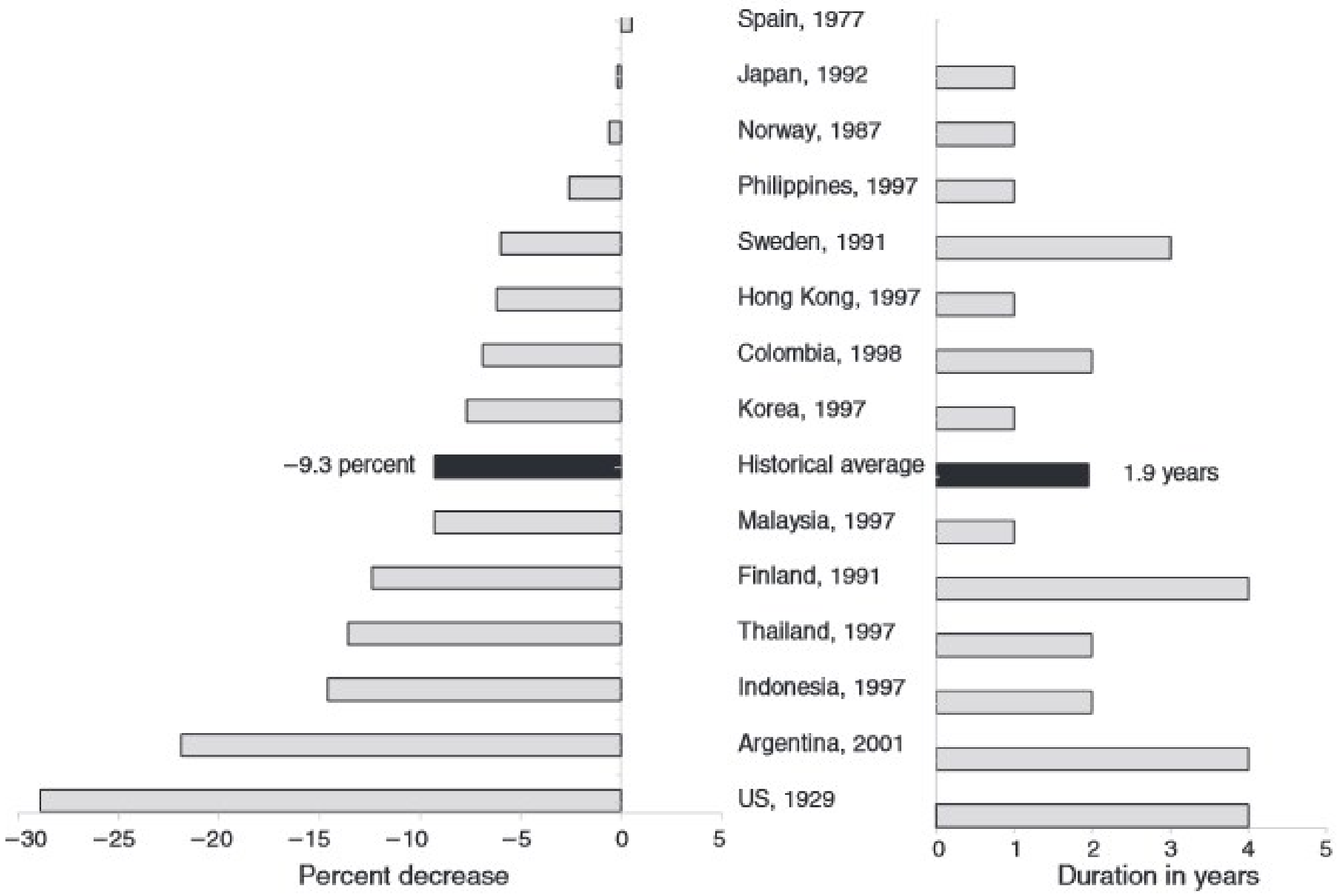

**Figure 2.1 Real GDP decline following a financial crisis, and duration of the decline in years. Source: Reinhart and Rogoff (2009)**

As shown in fig. 2.1, the two scholars record a persistent decline in real GDP for many countries affected by a financial crisis, with an average decline of 9.3% lasting 1.9 years. Although the consequences of the 2008 financial crisis and the consequent debt crisis are still hard to grasp, Rose and Spiegel (2009) report preliminary data, in line to the patterns found above.

| Country | % Change GDP, 2008 | % Change, Stock Market, 2009 |
|---|---|---|
| Iceland | -4.7 | -90.00 |
| Estonia | -2.8 | -63.00 |
| Latvia | -4.6 | -55.1 |
| Ireland | -2.8 | -7.8 |
| New Zealand | -0.9 | -37.4 |
| Italy | -0.6 | -49.5 |
| Denmark | -0.9 | -48.6 |
| Japan | -0.5 | -42.1 |

**Table 2.1 Change in GDP and average stock market values in selected countries. Source: Rose and Spiegel (2009)**

Indeed, financial crises impact the growth of societies, and any scholarly attempt that can help preventing them should be welcome.

Further historical evidence is presented in fig. 2.2, which shows a correlation between the occurrence of financial crises and the rise in unemployment. While the causal process leading from a crisis to unemployment may be contextual (Fallon and Lucas, 2002), OCSE studies point out that the 2008 financial crisis functioned as a catalyst that further increased unemployment – especially youth unemployment in Southern European countries (Choudhry, Marelli, and Signorelli, 2010).

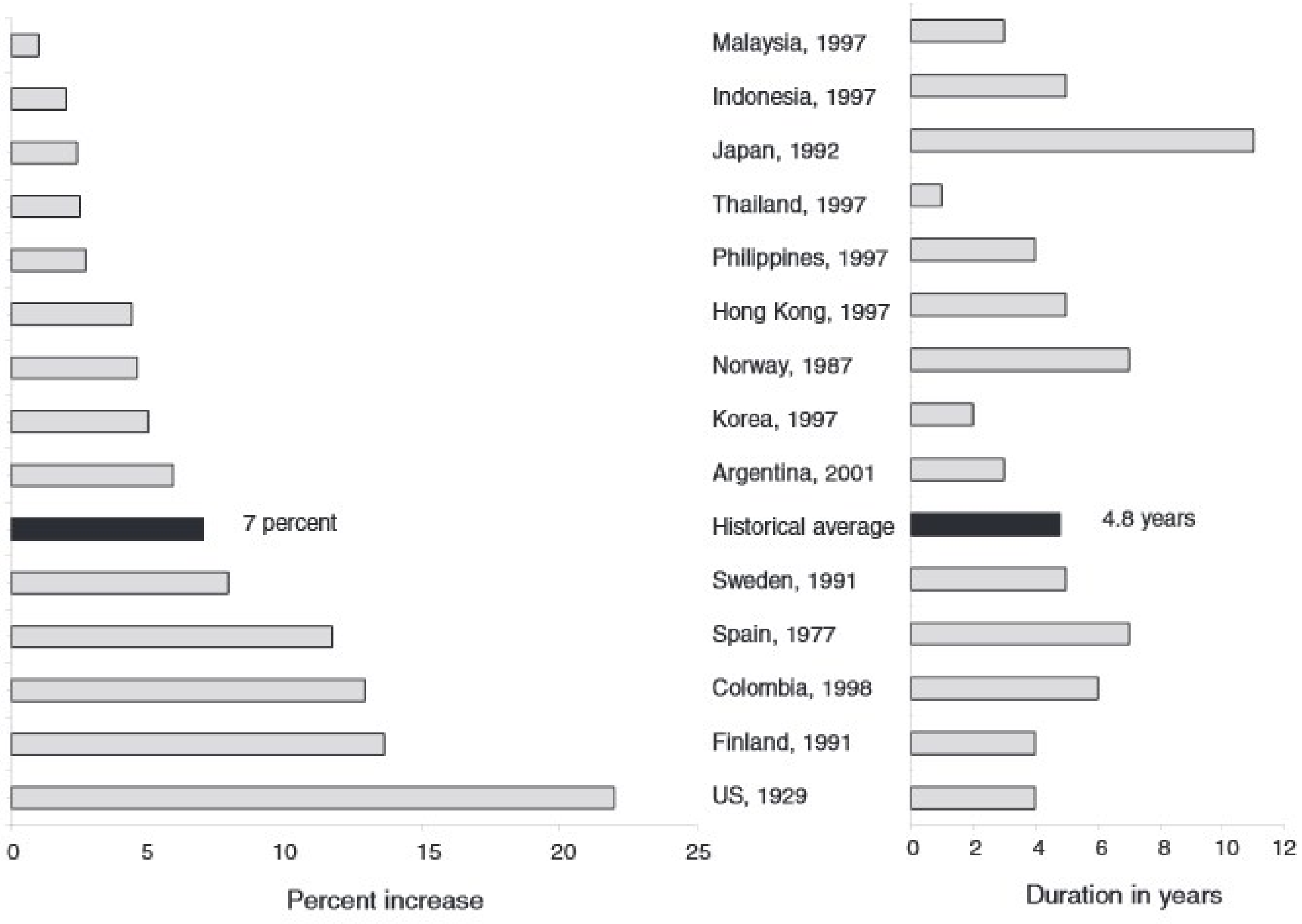

**Figure 2.2 Rise in unemployment following a financial crisis, and duration of the rise in years. Source: Reinhart and Rogoff (2009)**

It is worth noting that the relevance of the research is also highlighted by the lack of models and policy instruments that can provide insights to policymakers as to how adverse events spread. This will be made apparent by section 2.2.

## 2.2 Three approaches to systemic risk

The three strands of literature presented in this section attempt to yield policy instruments that can be used by the macroprudential policy maker. In doing so, they also attempt to answer the research question of this paper. Each line of contribution tries to face the problem using a different theoretical and methodological perspective. The first subsection focuses on the efforts coming from the regulators scholarship. They seek to understand how financial systems are affected by shocks using an indicator-based approach. The second subsection introduces some of the latest contributions originating from the application of DSGE models on financial systemic risk. These constitute the main reply of neoclassical economics to the argument that it has not been able to properly explain and account for modern-era financial crises. The third and last subsection explores the results of applications of complex network analysis and network contagion models on financial systems. It shows how depicting financial actors as components of a graph connected through interrelations can provide the most valuable insights onto how financial systems are affected by shocks. Each subsection also analyses their major flaws or weaknesses[1].

| Literature strand | Key research works | Strengths | Weaknesses |
|---|---|---|---|
| **Indicator-based approaches** | BCBS (2011)<br>BCBS (2013)<br>Masciantonio (2013) | • Easily implemented;<br>• Data availability | • Weighting is arbitrary;<br>• Disregard of systemic interrelations. |
| **DSGE modeling** | Tovar (2009)<br>De Walque, Pierrand and Rouabah (2010)<br>Gerali et al. (2010) | • Extensions of models already used by policy institutions;<br>• Integration with the simulations of the wider economic system | • Assumption of general equilibrium;<br>• Very small number of agents simulated;<br>• Very small number of time-steps considered. |
| **Complex network analysis** | Eisenberg and Noe (2001)<br>Battiston et al. (2012)<br>Roukny et al. (2013) | • Recognition of systemic interrelations;<br>• Capacity to quantify systemic effects of shocks and model their propagation | • Poor data availability;<br>• No uncertainty measures. |

**Table 2.2 Strengths and weaknesses of the literature considered**

### 2.2.1 *Indicator-based approaches*

Financial policy makers generally opt for frameworks that can be easily and robustly implemented, which require only public or readily accessible data, and are computable by non-scientific software (Adolfson et al., 2005). Because of such constraints model-based approaches are often not widely popular in several central banks and other regulatory bodies dealing with systemic risk.

A number of institutions confronted the issue by developing indicator-based approaches. The most relevant attempt is the one presented by the Basel Committee on Banking Supervision (BCBS), which identifies systemically important financial institutions (BCBS, 2011). The question of how an adverse event affects the whole financial system is then answered by stating the extent to which the institution hit by the initial shock is systemically important. The more it is, the more severe will be the threat to the whole system.

As mentioned in the introduction of this section, this line of literature has little explicit groundings on economic and financial theory; it is instead developed through several rounds of consultations with policy experts and empirical trials and evaluations[2] (FSB, 2009).

[1] Due to space constraints, the chapter does not, by any means, aspire to provide a truly comprehensive review of the relevant literature; it only seeks to guide through the essential scholarly work needed to contextualise the research. For more exhaustive accounts, see Galati and Moessner (2013).

| Category (weighting) | Individual indicator | Indicator weighting |
|---|---|---|
| **Cross-jurisdictional activity (20%)** | Cross-jurisdictional claims | 10% |
| | Cross-jurisdictional liabilities | 10% |
| **Size (20%)** | Total exposures as defined for use in the Basel III leverage ratio | 20% |
| **Interconnectedness (20%)** | Intra-financial system assets | 6.67% |
| | Intra-financial system liabilities | 6.67% |
| | Wholesale funding ratio | 6.67% |
| **Substitutability (20%)** | Assets under custody | 6.67% |
| | Payments cleared and settled through payment systems | 6.67% |
| | Values of underwritten transactions in debt and equity markets | 6.67% |
| **Complexity (20%)** | OTC derivatives notional value | 6.67% |
| | Level 3 assets | 6.67% |
| | Trading book value and Available for Sale value | 6.67% |

**Table 2. 2 Indicator-based measurement approach to identify systemically important financial institutions. Source: BCBS (2011)**

The Basel Committee's indicator conglomerates five distinct sets of categories, each arithmetically contributing to its value equally. They are, namely, cross-jurisdictional activity, size, interconnectedness, substitutability, complexity. As shown in table 2.3, most of the categories feature two or three operationalising measures.

Cross-jurisdictional activity indicates the importance of banks' activities outside of their home territories (Weistroffer and Speyer, 2011). A bank is deemed to be more systemically important if its activities are distributed across the globe. Size is another factor that, intuitively, might affect the importance of a bank. If a bank possesses high levels of total exposure, it is more likely to generate widespread distress in financial and economic systems. Interconnectedness relates to how a financial institution interacts with the rest of the system. According to the Basel Committee, the more an institution is connected to others, the higher is its systemic impact in case of failure or distress. Substitutability refers to how easily services of a financial institution can be replaced with another's. It is likely that the failure of a less substitutable bank would affect the health of the system more heavily. A negative correlation, thus, between substitutability and systemic importance exists. Finally, complexity regards the intricacy and ramifications of a bank's organisational structure and operations. Higher complexity would mean that the resolution of a default bank incurs in higher costs and negative externalities.

[2] Admittedly, the judgment of the policy experts that participated to the construction of the indicator is informed by theory, but a clear theoretical is neither declared nor discernible in technical documents (BCBS, 2011, 2013).

BCBS (2011) is not the sole body advancing indicator-based approaches to systemic risk. In fact, a number of studies followed its example to derive alternative indicators. All of them, nonetheless, tend to maintain the same categories of analysis, challenging only their operationalisations and proposing alternative sources of data or different calculation methods. A recent instance is Masciantonio (2013), which adapts BCBS' methodology such that only publicly available data is needed to construct the index. A further example is Braemer and Gischer (2011), which, in a similar way, proposes an indicator-based measurement of systemic banks and their impact that adopts only already-existing indices for each category. The approach has also been used to identify systemically important US non-banks financial institutions (FSOC, 2011) and insurance companies (IAIS, 2013).

Indicator-based measurements feature at least two critical shortcomings, respectively being methodological and conceptual in nature. These, arguably, prevent this stream of literature to properly answer the research question, thereby leaving room for novel, better approaches.
The first weakness is the underlying assumption that every category characterising the systemic importance of a financial institution is equally relevant. Even presuming that each of them is actually relevant, there is absolutely no a priori reason to justify equal weights. This only apparently naïve methodological decision is most probably dictated by the current uncertainty that still pervades the causal processes behind systemic risk and failures (Schwarcz, 2008). Some studies – such as Stern and Feldman (2004), Zhou (2010), Hart and Zingales (2011) – point towards the significance of financial institutions' size and complexity, whereas others – such as Drehman and Tarashev (2011), Dungey et al. (2012) – to their interconnectedness and international presence. To the knowledge of the author, currently no study or sensitivity test has been specifically designed to shed light on what indicators' categories contribute the most to the systemic stability of the financial system, and in what conditions. Nonetheless, this does not justify the choice of equally weighting the factors. Such a methodology can lead to biases in the estimation of the impact of shocks to the system, thereby causing erroneous policy decisions and detrimental consequences for the economy.
The second weakness relates to the conceptual implicit assumption that indicator-based approaches make: knowing the systemic importance of single financial institutions is enough to understand how a shock affects the whole financial system (BCBS, 2011). That is dangerously wrong. For instance, the failure of Northern Rock – a very systemically important bank – in late 2007 did not allow policymakers to assess the subsequent systemic consequences (Goldman-Pinkham and Yorulmazer, 2010). Even assuming that indicator-based approaches correctly identify the relevance of a bank, there is not enough information to comprehend how the shock propagates through other institutions, as it is not known how they are linked to each other (Minoiu and Reyes, 2009). That is why a network-based approach could be more suitable to the problem.

### *2.2.2 Dynamic stochastic general equilibrium models*

Dynamic stochastic general equilibrium models (DSGE) are increasingly used tools in economic and monetary policy making (Tovar, 2009). They are able to simulate the main actors of an economy – households, firms, governments – and their interactions by the computation of differential equations and utility functions which, according to neoclassical or neo-Keynesian economic theories, describe their behaviour (Sbordone et al., 2010). Because DSGE models allegedly have the faculty to capture – in a stylised way – the dynamics of an economy in its entirety, several central banks developed their own variants. Examples are the Bank of England (Harrison et al., 2005), the Bank of Canada (Moran, 1996) and the Central Bank of Chile (Chumacero and Hebbel, 2004).

Usually, DSGE models do not include financial institutions in their set of actors. If they do, they are assumed to be perfectly identical and frictionless, acting merely as intermediaries in the flow of money among all other economic agents (Tovar, 2009). This is certainly questionable, given the not-so-irrelevant frequency of economic shocks sparkled within the financial realm, such as the 1980-90s Japanese crisis, the 1980s Latin American crisis, the 1997 Asian crises, the dot-com bubble and the recent 2008 sub-prime crisis (Kindleberger and Aliber 2011).

Motivated by the gap in the literature, some scholars embarked onto the construction of novel DSGE models which would account for more sophisticated financial markets, thereby being able to understand how a financial system is impacted by shocks. Probably, the most representative instance of this line of research is De Walque, Pierrand and Rouabah (2010). The trio devises a DSGE model with a heterogeneous banking sector, that is, with banks whose portfolios and profit maximising behaviours are not identical. They also introduce endogenous default probabilities for both firms and banks, and allow for the simulation of banking regulations and liquidity injections in the interbank market. Such novelties permit the observation of the impact of external or internal shocks in the financial sectors, and would even enhance policy making through the experimentation of anti-systemic risk practices in the model. An analogous scholarly effort has been produced by Gerali et al. (2010). The paper introduces a DSGE model with financial frictions and with imperfectly competitive financial markets. The assumption of banking being in a market regime of perfect competition is relaxed, allowing for the existence monopolistic structures, as empirical studies suggest (De Bondt, 2005; Sorensen and Werner, 2006). Theoretically, such a feature is instrumental to simulate realistic shocks dynamics, as the size and market concentration of financial institution may well affect the health of the overall system.

A DSGE approach to systemic risk model has severe limitations in assessing the systemic effects of shocks in finance. Two of them are explored in this paragraph.

DSGE models equations are deeply rooted in neoclassical and, more recently, neo-Keynesian theories (Wickens, 2012). That means that any bias that affects such theories is also reflected in the computational models. The most critical assumption is that of general equilibrium: it is granted that there exists a steady state for the economy, towards which it tends to converge over time (Starr, 2011). The assumption is regarded as unrealistic. Closer empirical inspection suggests that economic and social processes are dynamic and in a state of evolvable change and multiple equilibria (Beinhocker, 2007; Hodgson and Knudsen, 2010). In fact, what characterises economies is its ever-changing face, e.g. new industries arising and others declining, changes in labour demands due to innovation (Arthur, 2006).

The practical implication of the general equilibrium assumption is an underestimation of the propagation of shocks within the system. Simulated shocks make the system diverge from the steady state, whereas the equations dictating the behaviour of financial actors will act in such a way as to let the system return it (Colander et al., 2009). This negative feedback is likely to reduce or nullify the positive feedback of the shock, acting as a bias.

The second weakness relates to the over-simplicity featured in DSGE models. The one in De Walque, Pierrand and Rouabah (2010) counts only two firms and two banks. Also Gerali's et al. (2010) model simulates only a very small number of financial institutions. As for February 2014, there are 157 banks incorporated in the UK alone (BoE, 2014). Without a realistic number of financial actors, it is not possible to obtain informed accounts on how shocks affect the financial system. The main reason is that with higher number of market participants the shock dynamics may substantially change qualitatively, a phenomenon known as emergence (Standish, 2001; Gregersen, 2002). The 2008 financial crisis came to being through the non-trivial interaction of hundreds of actors, not simply a handful (Taylor, 2009).

### *2.2.3 Network analysis*

The shortcomings of the two approaches explored above are mainly due to the disregard of the complexity of financial systems. Both lines of contributions experienced a discrete success in policy making because of the easiness in their elaborations and the compatibility with already-existing policy instruments – at the cost of gravely compromising the ability to understand the dynamics of credit crises. The costs far outweigh the benefits, and an alternative, more empirically-valid and elaborate approach is needed. Complex network analysis, arguably, is the good candidate.

As Haldane and May (2011) recognised, the interaction of financial actors gives rise to patterns of complex interactions and nonlinear dependencies, analogous to the ones found in natural and social ecosystems. This allows the application of complex systems analytical methods to the problem of study, with particular reference to the analysis of complex networks (Waldrop, 1992; Bogg and Geyer, 2007).

Network analysis is the study of the topology, structure and dynamics of systems whose component's relations can be mapped onto node-link graphs (Wasserman, 1994). Some early studies showed the importance of interbank relations with regards to systemic risk. Allen and Gale (2000) provided the theoretical foundations of how shocks can spread through a network of financial institutions, whereas Freixas and Parigi (1998), Eisenberg and Noe (1999) devised the very first financial shock contagion models, whose algorithm is introduced in Chapter 3. The latter work inspired the empirical application of network analysis in financial systems. Examples are Upper and Worms (2004) for the German financial market, and Boss and Helsinger (2004) for the Austrian one.

The lack of appropriate data slowed down the advancement of this line of research. Information about interbank linkages, such as short and long-term lending and portfolio commonalities, is often lacking (Upper, 2007). In the last few years, though, scholars have produced seminal work with the potential of properly answering the research question, thanks to new data gathered by central banks and the appearance of novel statistical methods to treat partial information (Servedio, Caldarelli and Butta, 2004).

Two research papers are of key interest. The first is Battiston et al. (2012), which develops a new measure of systemic risk known as DebtRank. DebtRank is a recursive algorithm that is able to provide estimates of the total systemic loss caused by the failure or distress of one or more banks. In other words, it simulates how a market shock would spread across financial institutions, quantifying the final damage of the cascade effect. The authors apply the instrument to the network of FED emergency loans (comprising a large number of international banks), and identify what banks – according to available data – would cause the most adverse effects on the whole network.

The second research has been produced by Roukny et al. (2013), who explore the stability of different network topologies against default cascades. More specifically, the authors simulate default cascade dynamics in banking networks under several market conditions, such as market illiquidity and sudden capital ratios changes. The topologies of the simulated networks are scale-free or random. Their results show that no architecture can always be preferred to another, yet their methodology provides a valuable starting point to assess how market shocks propagate among financial networks. Similar conclusions are found in Huang et al. (2013), who simulate default cascades caused by market price shocks.

Also these works are not immune from complications, which give rise to gaps that the research intends to fill. The most critical of which is still the poor availability of interbank data. Battiston et al. (2012) relies on a chunky dataset, where only partial data is available and most probably not representing the totality of the interbank financial network. In a similar vein, other empirical studies such as Iori et al. (2008) make use of data referring to overnight or short-term lending in national markets. This approach may not allow inferences to be valid on the populations of interest, such as networks of international banks and networks of longer-term financial relationships. Ultimately, the risk of missing crucial mechanisms of how shocks propagate is nontrivial. For this very reason, it is needed a method that would allow researchers to reconstruct networks from incomplete data, and would couple it with a financial contagion model. This strategy would also permit the integration of uncertainty measures into the investigation of systemic risk. Chapter 3 will discuss this very point, among other things.

## 3. The model: methodology and design

The model described in this section attempts to quantitatively understand how financial systems are affected by adverse events, provided that incomplete data is available with regards to how banks dynamically affect each other. It does so by simulating shocks on an ensemble of financial networks reconstructed from partial real information[3]. The Debtrank and default cascade algorithms are used (and their performance compared) to quantify the severity of the shocks caused by each bank in distress. Uncertainty about the actual impact of the shock is highlighted by the use of confidence intervals for the resulting output of the algorithms. The next subsection will describe the key concepts of the model, together with their operationalisations. Subsections 3.2 and 3.3 will outline the contents of the main two modules of the model, respectively dealing with networks estimation and contagion simulation.

### 3.1 Concepts and operationalisation

As summarised in table 3.1 below, the research makes use of four key concepts derived from network science and finance.

| Concept | Operationalisation |
|---|---|
| **Network** | Interbank market network |
| **Nodes** | 227 largest (by total asset value) EU15 financial institutions indexed in stock markets |
| **Node attributes** | Financial institutions' total assets, market capitalisation, total interbank assets, total interbank liabilities |
| **Edges** | Financial institutions' exposure to each other |

**Table 3.1 Concepts and respective operationalisations**

Of utmost importance is to define the concept of network, nodes, and edges. A network (or graph) is a collection of entities, known as nodes, and a collection of relationships between nodes, known as edges or links (Wasserman, 1994). Analytically, a network can be simply represented as a set containing all nodes and edges:

$$\mathrm{G} = \{N, E\}$$

$$N = \{n_1, n_2, \dots, n_g\}$$

$$E = \{e_1, e_2, \dots, e_k\}$$

**Equation 3.1 Analytical representation of a network**

[3] Despite the use of real world input data, the inferences of the model cannot by any means be considered of empirical value. The data is simply used in order to construct networks with realistic topologies.

Where $n_g$ is a single node in the network, and $e_k$ is an edge between two nodes. An alternative way is to represent a network using an adjacency square matrix, with dimensions equal to the total number of nodes in the network. At the intersections between the rows and columns of the adjacency matrix edge values are reported.

Each node can store attributes, qualitative or quantitative properties that provide specific information about the nature of the nodes (Brandes and Wagner, 2004). Edges can be undirected or directed. If directed, the kind of relationship represented is asymmetric. Furthermore, edges can also be weighted with a value explicating the intensity of the relationship (Palla et al., 2007). This research is focused on directed weighted edges, which can be analytically represented by a simple vector containing the source node, the end node, and the value of the weight:

$$\overrightarrow{e_k} = < n_i, n_j, w >$$

**Equation 3.2 Analytical representation of a directed weighted edge**

In this model, a network represents a collection of synthetic interbank lending snapshots of financial institutions estimated for the 227 financial institutions with largest total asset value, located in the EU15[4] and indexed in a stock market. Each snapshot would refer to the synthetic market exchanges registered over a month. More specifically:

(i) Nodes constitute financial institutions classified as commercial and investment banks, that is, focused on household lending or on the provision of capital to large business projects and securities (Haberman, 1987). Each node is characterised by four scalar attributes: its market capitalisation, representing the robustness of the bank (Ediz et al., 1998), its total asset value; and the value of their total interbank assets and liabilities. The data is extracted from the database *Bankscope*. It refers to the end of the fiscal year 2007, and is used to estimate the synthetic networks adopted in the experiment. By doing so, it is assured that the synthetic network topologies and dynamics approach those of real financial networks. Subsection 3.2 better elaborates this point. As shown in fig. 3.1, the frequency distributions of the data clearly show power law features, in which few actors in the system have values of order of magnitude greater than the majority of all other actors.

[4] Austria, Belgium, Denmark, Finland, France, Germany, Greece, Ireland, Italy, Luxembourg, Netherlands, Portugal, Spain, Sweden and United Kingdom.

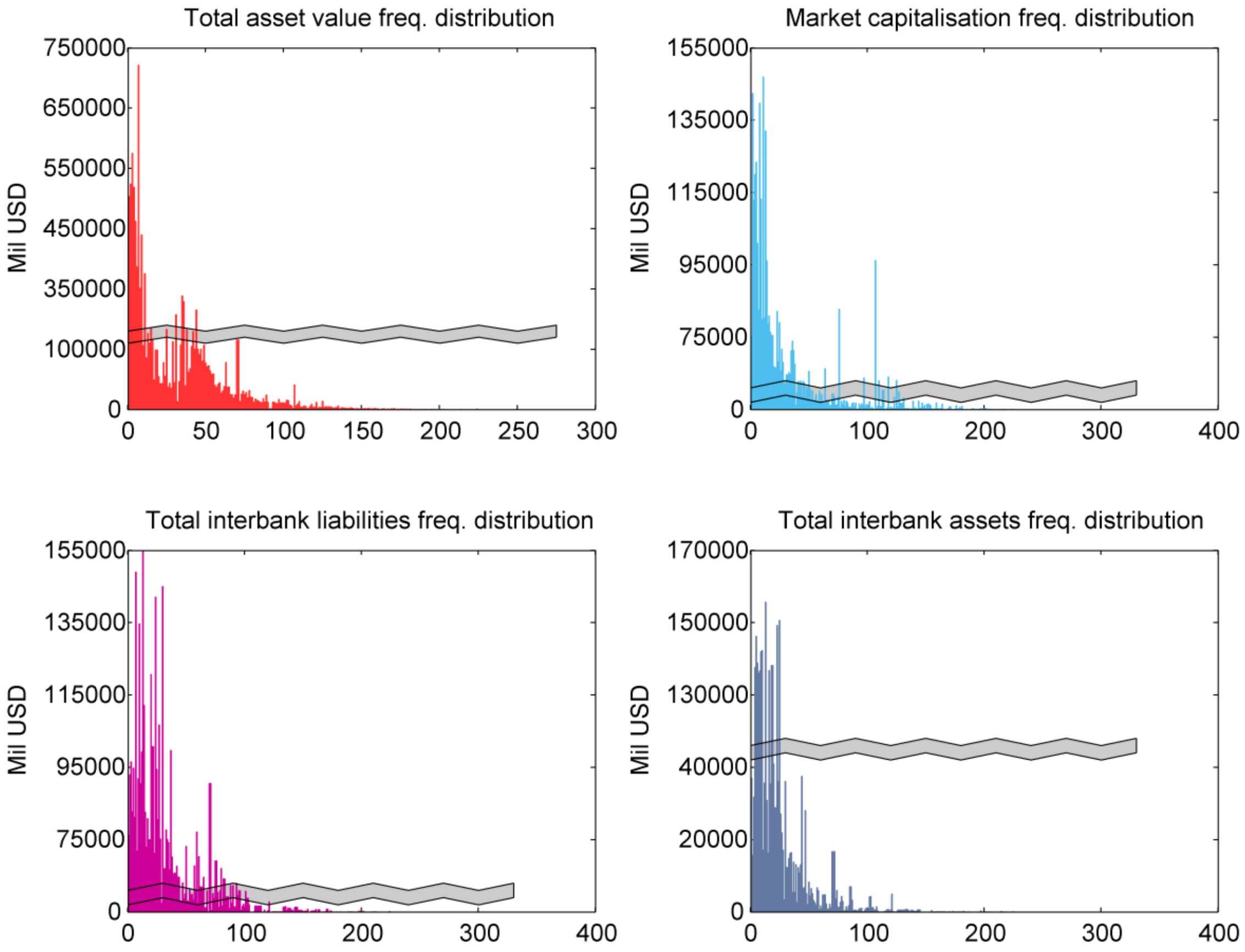

**Figure 3.1 Frequency distribution of banking data**

(ii) Interbank connections are operationalised by the amount of not yet claimed lending between financial institutions over a pre-determined period of time. The lending represents a directed edge from the borrower to the lender, with weight equal to the amount borrowed. Such quantities are calculated using the algorithm described in the following paragraphs.

## 3.2 The network estimation module

The network estimation module is based on the fitness bootstrapping method presented in Musmeci et al. (2013). As in De Masi, Iori and Caldarelli (2006), the method assumes that the probability of any two nodes being linked is proportional to a latent scalar value, named the *fitness* of the node. Analytically:

$$p_{ij} = \frac{x_i x_j}{1 + x_i x_j}$$

**Equation 3.3 Probability of the existence of an edge between node *i* and *j* as a function of their fitness**

Where $p_{ij}$ is the probability of nodes $i$ and $j$ being linked, and $x_i$ and $x_j$ their respective fitness values. Because $x$ is latent, it is assumed the existence of an observable proxy of the fitness value, $y$, evaluated from $x$ though an unknown universal parameter $z$:

$$\sqrt{z}y_i = x_i$$

**Equation 3.4 Relation between the latent fitness and its proxy**

Therefore, eq. 3.3 becomes:

$$p_{ij} = \frac{zy_iy_j}{1 + zy_iy_j}$$

**Equation 3.5 Probability of the existence of an edge between node i and j as a function of their fitness proxy values**

Eq. 3.5 can be used to estimate a network given the knowledge of only some or of none of the edges, i.e. of a sub-network. This is possible through the estimation of $z$ in the following equality:

$$\sum_i k_i \equiv \sum_i \sum_{j \neq i} p_{ij}$$

**Equation 3.6 The relationship between the sub-network sum of degrees and the sum of edge probabilities**

Where $k_i$ is the degree of each node in the sub-network – i.e. the number of edges it possesses. In the case of a total lack of knowledge of a sub-network, $\sum_i k_i$ can be substituted with a pre-determined amount of total links. Once $z$ is found, it is possible to estimate the edge probabilities for each pair of nodes, and estimate a network ensemble.

In financial networks, it has been demonstrated that the value of total assets of a bank performs well as a fitness proxy variable. Musmeci et al. (2013) have found that, given the availability of total asset information for each bank and information about 7% of the edges, it is possible to reconstruct an entire network with same topology and a 7-10% margin of error, meaning that around a tenth of the estimated edges or non-edges do not coincide with the real network. The module follows this best practice in order to obtain realistic networks with pre-specified densities.

Once estimated the networks' topologies, the weight of each edge are to be estimated. This is not a trivial problem, as the weight between any two nodes must not be greater that the source node's total interbank asset and the end node's total interbank liabilities. Analytically, each weight has two constraints:

$$0 < w_{ij} < \sum_i A_i^{IB}$$

$$0 < w_{ij} < \sum_j L_i^{IB}$$

**Equation 3.7 Constraints for network weights**

The problem is here solved with the heuristic explicated in fig. 3.2. The algorithm cycles through each node *i* in the network, allowing it to lend a small portion[5] of its total interbank assets to its successor nodes, say *j*. The iteration on a lender node stops when the total amount lent reaches $\sum_i A_i^{IB}$, while the iteration on a borrower node ends when $\sum_j L_i^{IB}$ is reached. In order to diminish computing time, the algorithm globally stops when the entire networks has been iterated 500 times[6].

[5] The amount lent at each iteration is 0.01 mil USD.

[6] The algorithm allows constraints to be respected, but the total sum of the resulting weights will not obviously equal the real-world interbank liabilities and assets value. This is due to the fact that real financial institutions have interbank exposures to institutions not included in this model.

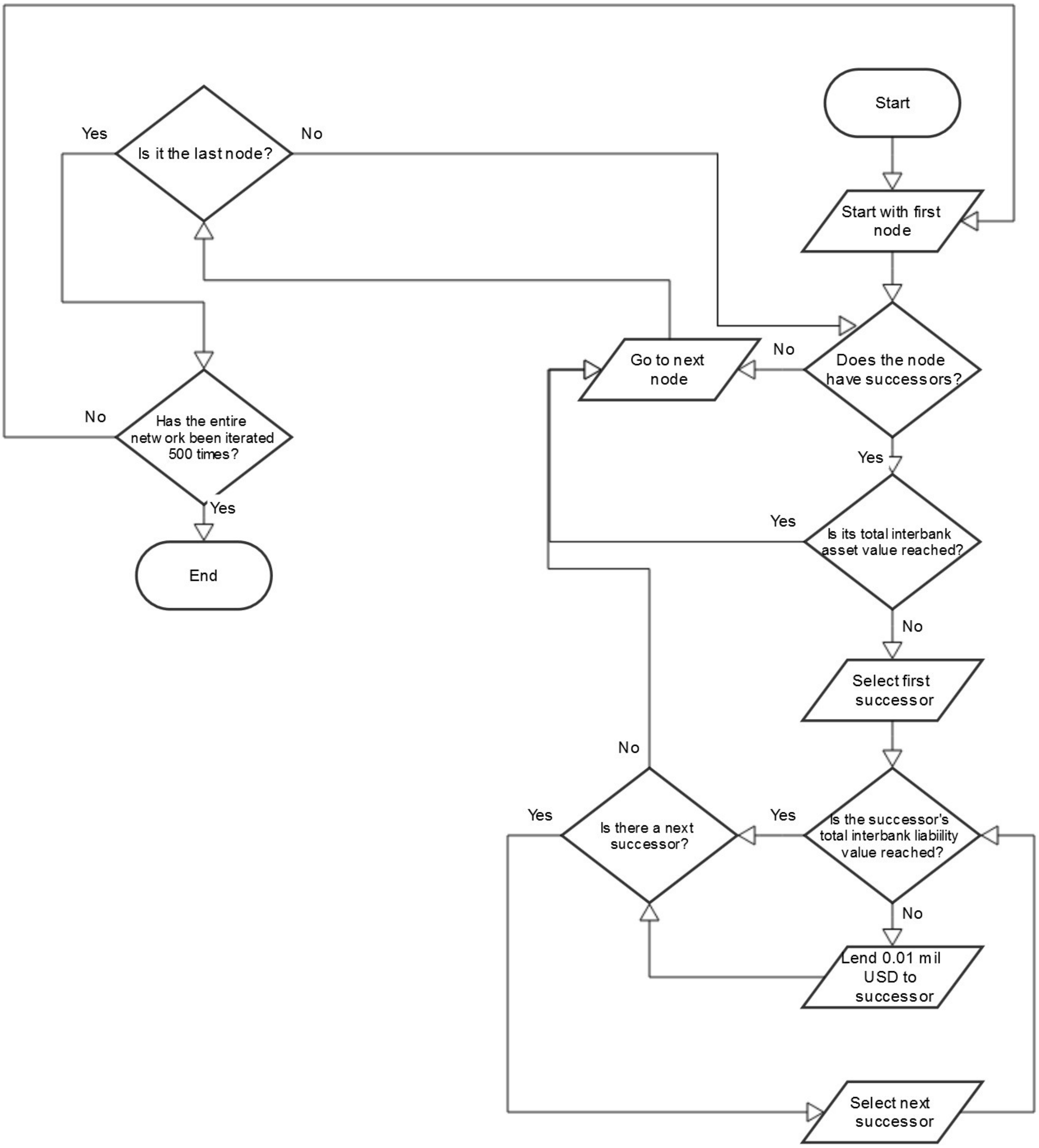

**Figure 3.2 Flowchart of the algorithm assigning weights to the estimated links**

The final flow of the module is presented in fig. 3.3. For each snapshot desired in the experiment, an ensemble containing 50 networks is estimated. This, as it will be shown in Chapter 4, will allow the evaluation of uncertainty intervals in the experiment.

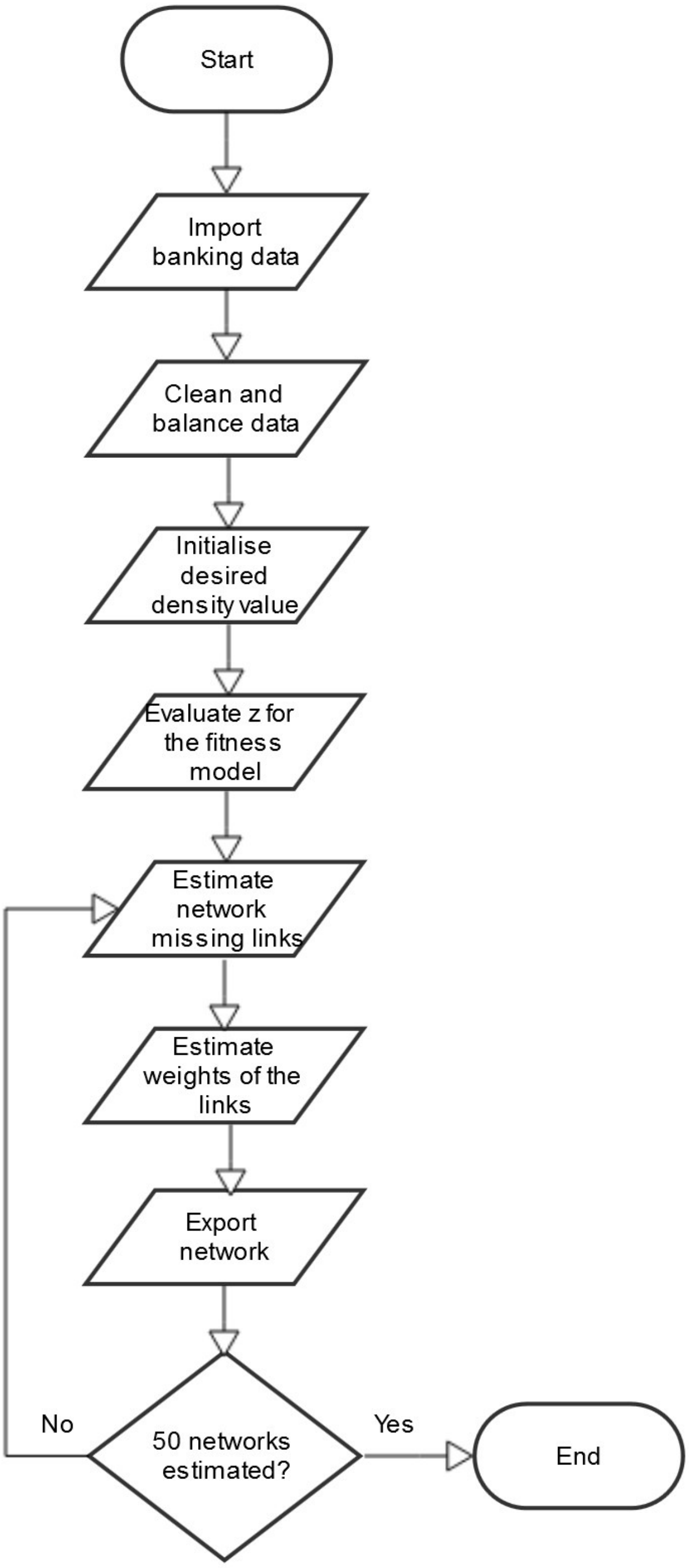

**Figure 3.3 Flowchart of the network estimation module**

### 3.3 The contagion module

The contagion module adopts DebtRank and the default contagion algorithm (Battiston et al., 2012) in order to model the impact of financial shocks in the networks reconstructed through the previous module. Both algorithms are recursive methods suggested to determine the systemic relevance of nodes in financial networks. They measure the fraction of the total economic value in the network that is potentially affected by one or more nodes (Thurner and Poledna, 2013). Analytically, they work as a dynamic epidemiological process.

Two inputs are needed by DebtRank. The first is $W$, the impact matrix, with same dimensions as the network's adjacency matrix. In this work, each element of $W$ is evaluated by dividing the liabilities of the borrower by the market capitalisation of the lender:

$$W_{ij} = min[1, \frac{L_{ij}}{C_j}]$$

**Equation 3.8 Impact of node *j* (the borrower) to node *i* (the lender)**

Where $L_{ij}$ is the amount borrowed by $i$ from $j$, and $C_j$ is $j$'s capitalisation. This measure provides an indicator of the vulnerability of a lender with respect to the borrower. If, for instance, the lender's exposure to a borrower has value 0.2, it implies that the lender would lose an amount of assets equal to 20% of the value of its capital if the borrower failed. This would put the lender in distress, but it would not cause it to default, as there is still capital that can cope with the loss.

The second input is a vector indicating the initial shocks to be simulated, $\psi$. $\psi$ has the same length as the total number of nodes in the networks; each element can take a value between 0 and 1, where 0 means no shock and 1 implies the default of the bank.

Once the inputs are initialised, two state variables are assigned to each node in the network: a continuous variable $h_i \in [0,1]$, and a discrete variable $s_i \in \{U, D, I\}$. At time 1, their state is:

$$\begin{aligned} h_i(1) &= \psi_n, \qquad \forall i \in S_f \\ h_i(1) &= \psi_0, \qquad \forall i \notin S_f \\ s_i(1) &= D, \qquad \forall i \in S_f \\ s_i(1) &= U, \qquad \forall i \notin S_f \end{aligned}$$

**Equation 3.9 Initial conditions for state variable *h* and *s***

Where $S_f$ is the set of nodes in distress at time 1, and $\psi_n$ and $\psi_0$ are respectively non-zero and zero values in . The contagion dynamics is described by eq. 3.10 and 3.11:

$$h_i(t) = min\left[1, h_i(t-1) + \sum_{j|s_j(t-1)=D} W_{ji}h_j(t-1)\right]$$

**Equation 3.10 Dynamics of variable *h for DebtRank***

$$s_i(t) = \begin{cases} D \quad if\ h_i(t) > 0, s_i(t-1) \neq I \\ I \qquad\qquad if\ s_i(t-1) = D \\ s_i(t-1) \qquad\qquad otherwise \end{cases}$$

**Equation 3.11 Dynamics of variable *s***

Finally, the impact $R$ of nodes in set $S_f$ is simply the difference between the total impact at the end of the cascade at time $T$ and the initial shock, multiplied by the nodes' economic value $v$. In this application $v$ is represented by the banks' market capitalisation values:

$$R = \sum_i h_i(T)v_i - \sum_i h_i(1)\,v_i$$

**Equation 3.12 Evaluation of the systemic impact**

As specified in eq. 3.11, the impact can have currency value. If DebtRank is evaluated for each financial institution (for instance, by simulating the default of each), the resulting $Rs$ can be normalised so as to obtain a ranking of their systemic importance.

The default cascade algorithm can be described by the same mathematics outlined above, with two crucial differences. First, the shock values in $\psi$ can only be ones (as it only simulates complete defaults); second, the shock propagates if and only if an affected node collapses. Eq. 3.13 changes accordingly:

$$s_i(t) = \begin{cases} D \quad if\ h_i(t) > 1, s_i(t-1) \neq I \\ I \qquad\qquad if\ s_i(t-1) = D \\ s_i(t-1) \qquad\qquad otherwise \end{cases}$$

**Equation 3.13 Dynamics of variable *s* the default cascade**

Clearly, there are limitations to this approach, which become evident thanks to the results of the experiment. Fig. 3.4 summarises the flow of the contagion module as it will be run in the experiment. Because several time snapshots with synthetic networks are evaluated, the module cycles through them.

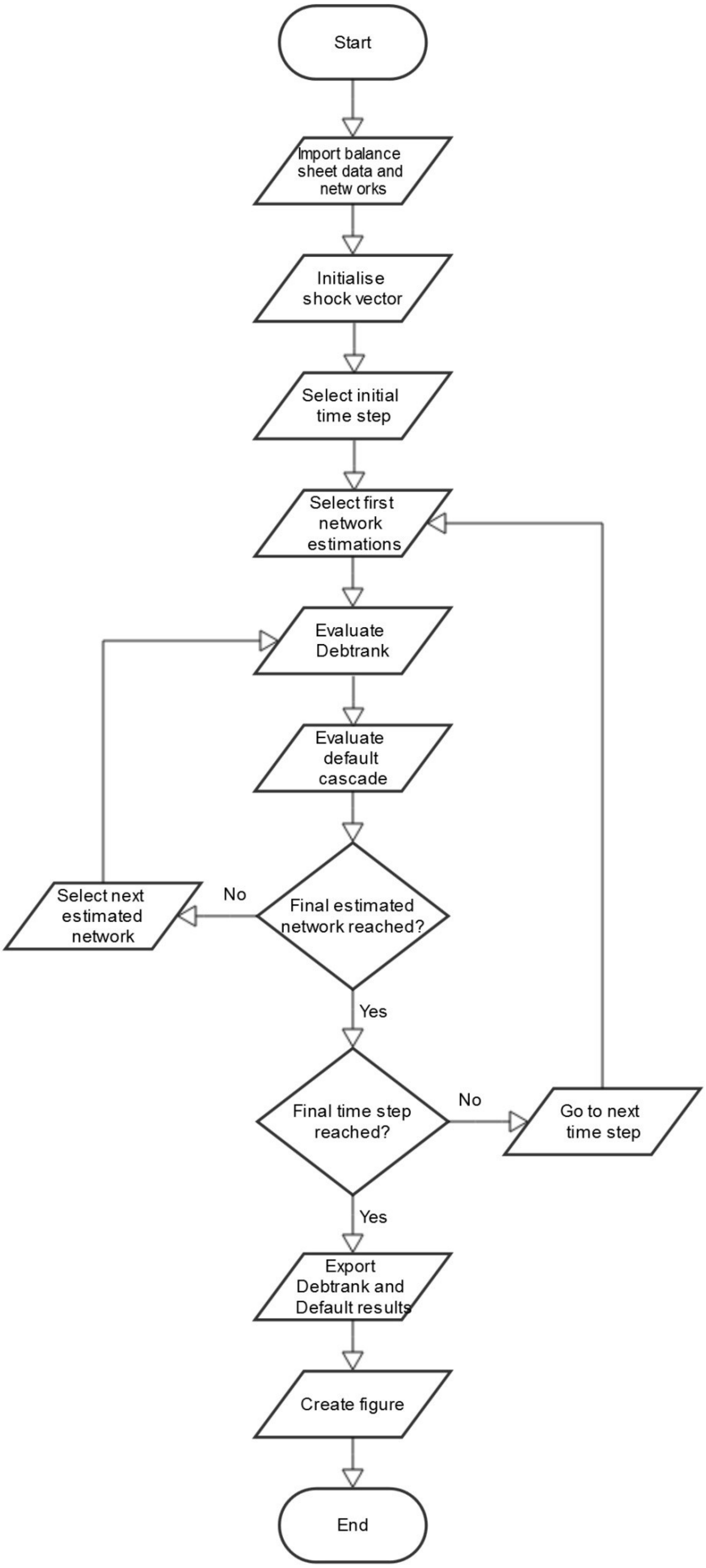

**Figure 3.4 Flowchart of the contagion module**

## 4. The model: Implementation and results

This research contends that more light can be shed to the question of how financial systems are affected by adverse events by the use of the DebtRank algorithm, augmented with uncertainty intervals. The synthetic experiment of this section aims at testing its performance against another measure based on the complex networks analytic framework, namely the default contagion algorithm. More specifically, it tests hypothesis $H_0$:

*$H_0$: The Debtrank algorithm is able to capture losses due to adverse events in an interbank network sooner than the default contagion algorithm.*

In the experiment, losses are measured as the fraction of the total market capitalisation of the network that is lost after the shock. Shocks are the failure of one of the financial institutions. By 'sooner', instead, it is meant the capacity of identifying systemic losses in a balance sheet setting where the other algorithm still can't.

This chapter is divided into two sections. The first will present the experiment setting, and the three synthetic network ensembles on which the hypothesis is tested. The second will demonstrate the result of the analysis and conclude that DebtRank indeed outperforms the default contagion algorithm in the task of identifying systemic losses.

### 4.1 Network ensembles and experiment settings

By the use of the first module of the model, three network ensembles are reconstructed from the real-world data shown in the previous chapter. Each ensemble consists of 50 networks, enough to build confidence interval of the systemic losses estimates. Such network ensembles possess identical node attributes (that is, balance sheet data), but will differ in density distribution, in order to allow the testing of the financial contagion algorithms at different topological settings. More specifically, each ensemble respectively features about 1500, 3000, and 5000 edges. Summary statistics are shown in table 4.1 below.

| $\sum k$ | Mean | St. Dev. | Min | Max |
|---|---|---|---|---|
| **1500** | 1502.20 | 28.85 | 1445 | 1568 |
| **3000** | 3003.10 | 35.06 | 2939 | 3102 |
| **5000** | 4995.90 | 37.25 | 4912 | 5073 |

**Table 4.1 Summary statistics of the density distributions for the three ensembles used in the experiment**

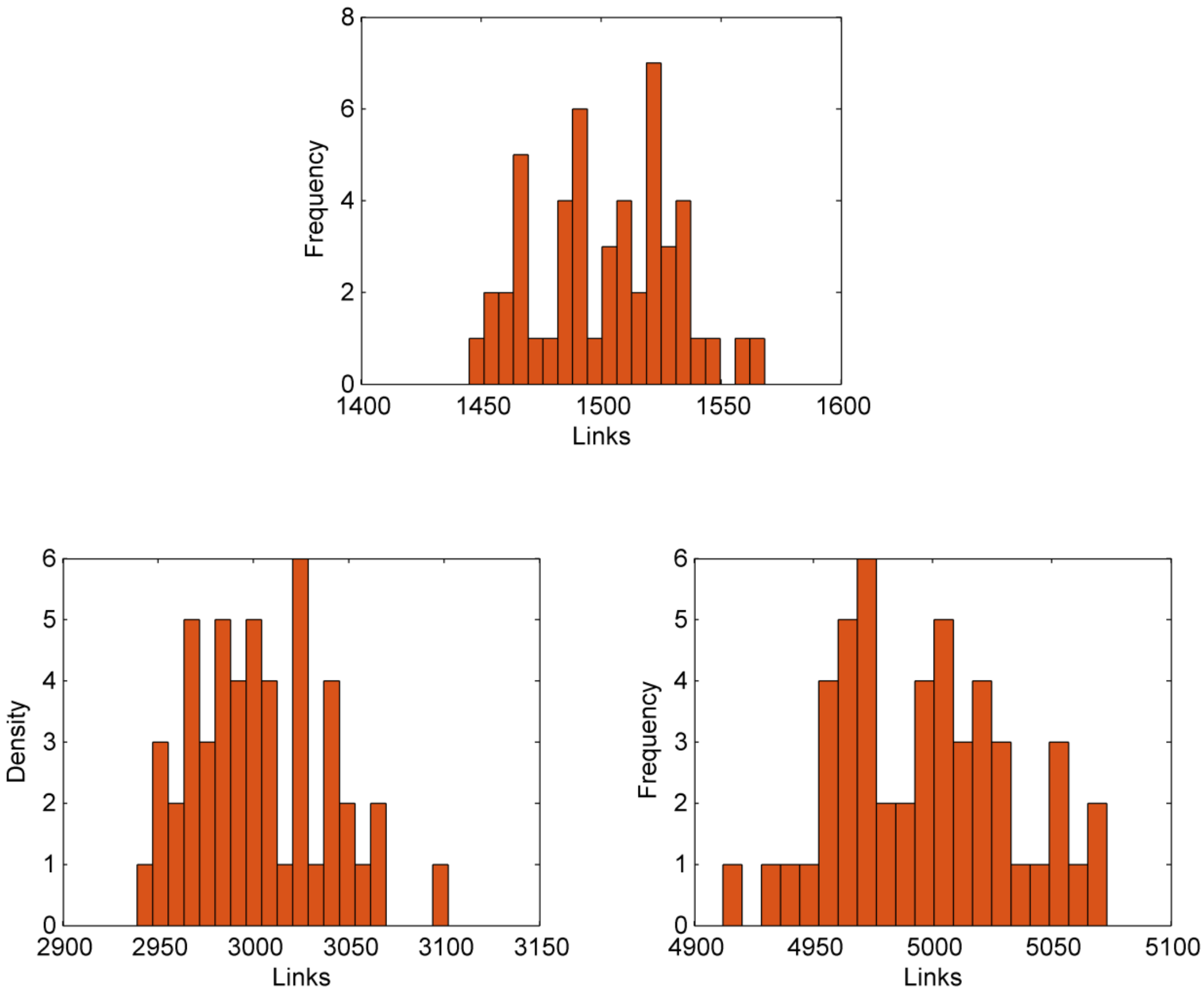

**Figure 4.1 Frequency distributions of the density in each ensemble**

The distribution of the total sum of links for each ensemble is displayed in fig. 4.1 above. The module estimates collection of networks with densities that oscillate around the inputted $\sum k$ in eq. 3.6. Although it cannot be stated conclusively because of the relatively small number of networks in each sample, it is likely that the asymptotic form of the frequency distributions above is normal. Further investigation will have to ascertain the contention[7].

Because of the stochastic nature of the estimation method, several possible link configurations for each node are explored. This engenders the possibility to evaluate a distribution of the impact of the systemic loss caused by a node, as calculated by the two algorithms. This very opportunity enables the construction of the confidence intervals which will be seen in the next section. The node-degree distributions of the networks resemble the ones shown in fig. 4.2. As the density increases, the power law form of the distributions appear to flatten out.

[7] The statistical properties of this network estimation method have not yet been explored, to the knowledge of the author.

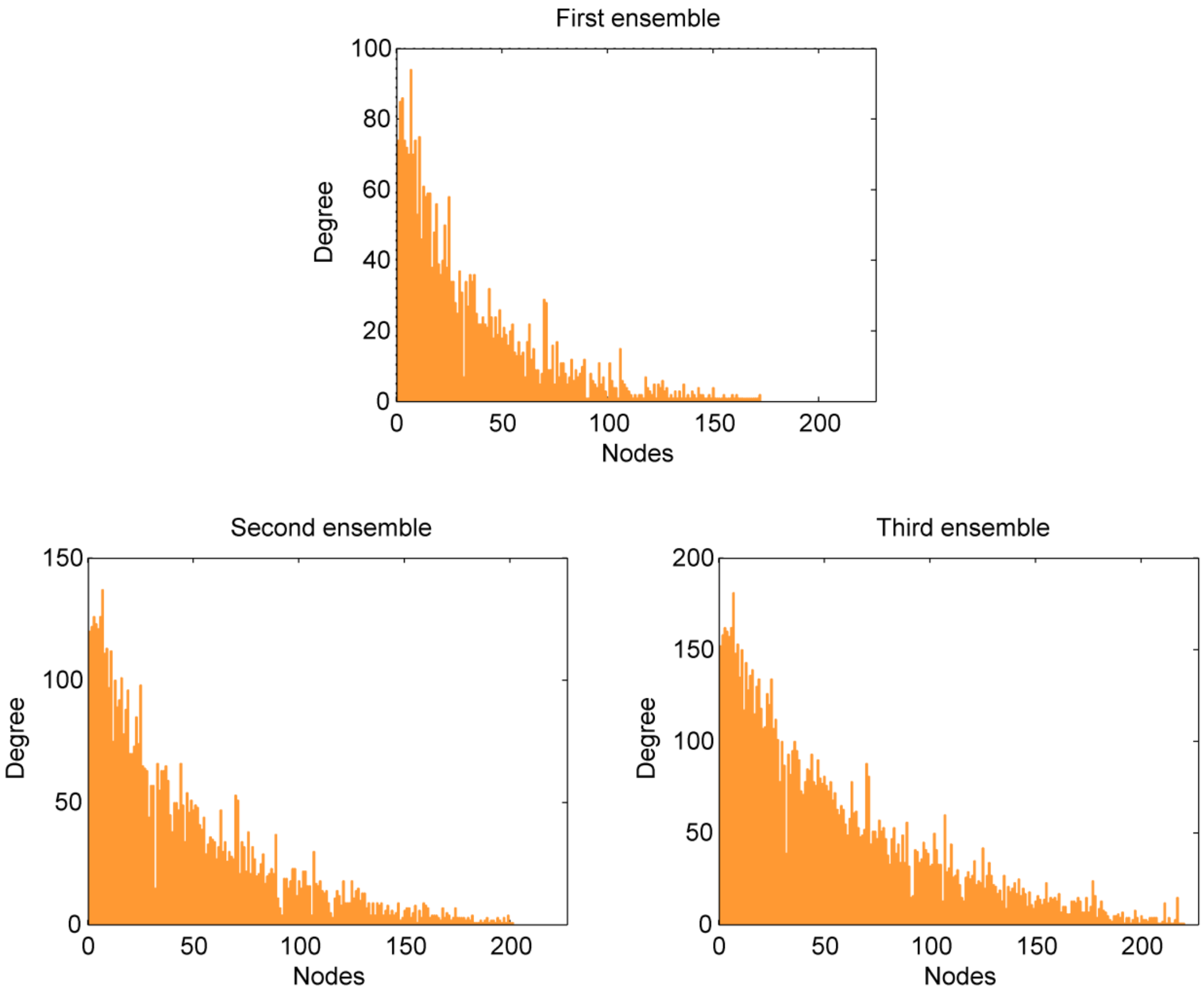

**Figure 4.2 Node-Degree distribution in example network for each ensemble**

The experiment tests the performance of the contagion algorithms by building an artificial time series of network snapshots with gradually worse systemic conditions, which will render each of the banks in the network extremely systematically important. In other words, by the end of the time series, the failure of each bank would cause almost the entire financial network to collapse. Operationally, this can be done in two ways: by gradually increasing the amount of interbank lending, or by gradually decreasing the market capitalisation of the banks. In both case the relative dependency on each bank to another increases, thereby causing greater distress in case of a shock. In this experiment, it has been decided to opt for the latter method. Ten time steps are constructed. At each step, the market capitalisation of each bank becomes on third of what the value was in the previous step. Analytically:

$$C_t^i = 0.3 * C_{t-1}^i, \quad t = 0, \dots, 10$$

**Equation 4.1 Changes in market capitalisation at each time step**

Where $C_t^i$ is the market capitalisation of node $i$ at time step t. Apart from the change in capitalisation, all conditions remain the same. The networks at the each time step feature same topologies and balance sheet attributes.

The experiment works as follows. For each time step, each node is assumed to fail (i.e. each bank to collapse), and the proportion of the losses of the system caused by the failure are evaluated with both DebtRank and the default contagion algorithm. The losses shown are the arithmetic average across the ensemble, with maximum and minimum variables being respectively the upper and lower uncertainty bounds. Because of space constraints, the outputs of the experiment are shown only with regards to the first five largest nodes by total asset value[8].

### 4.2 Results

Figures 4.3, 4.4 and 4.5 respectively show values of DebtRank (left columns) and default contagion (right columns) for the three ensembles. The horizontal axes are time steps, while the vertical ones represent the proportion of total market capitalisation in the system which has been burnt as a consequence of the failure of the bank. It is evident that DebtRank is able to capture the increasing systemic importance of the collapsed banks in a more continuous and sensible way than the default contagion algorithm. $H_0$, therefore, is not rejected. From a deeper scrutiny of the results, the following observations can be made:

(i) Both algorithms converge at the final time steps, when the financial institution is already capable of letting almost the network collapse. This shows the value of the default contagion method in capturing the systemic importance of a bank and, therefore, the fragility of the system with respect to it. Nevertheless, it does it too late, when the losses are already at great magnitudes.

(ii) When the default contagion algorithm still registers zero systemic impact, DebtRank features increasingly steeper values, thereby acting as an early warning signal of the later total collapse. It is possible to note increasing nonzero values of the impact at time step 4 for all nodes, for instance.

(iii) By increasing the density of the networks in the ensemble, on average the impact of the shocks becomes slightly greater. Surely, a greater variation in density is needed to inspect further this observation, yet it can already be noticed that an increase in the interbank exposures is able to augment the rate at which impact cascades occur.

(iv) The uncertainty intervals of the default cascade algorithm are extremely large, while the ones of DebtRank are narrower, with a convergence towards the final time steps. This suggests a greater ability of this variant of DebtRank to capture the systemic importance of a bank despite the lack of knowledge of its exact links configuration, a property that is extremely valuable in a context where the lack of fine-grained data is persistent.

[8] The entire dataset with the results, though, is available for inspection.

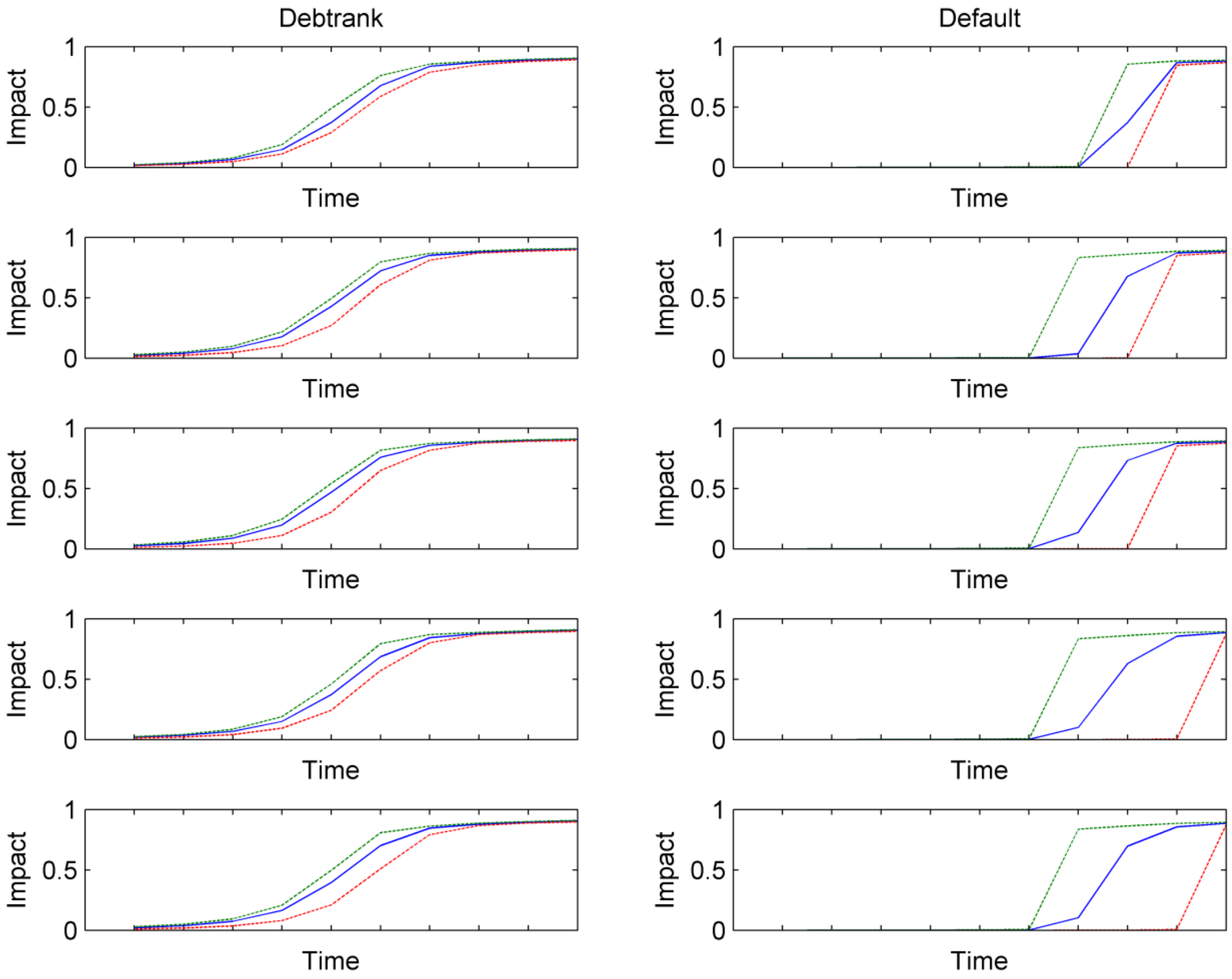

**Figure 4.3 Evolution of Debtrank and default contagion for the five largest banks for the ensemble with $\sum k$ = 1500**

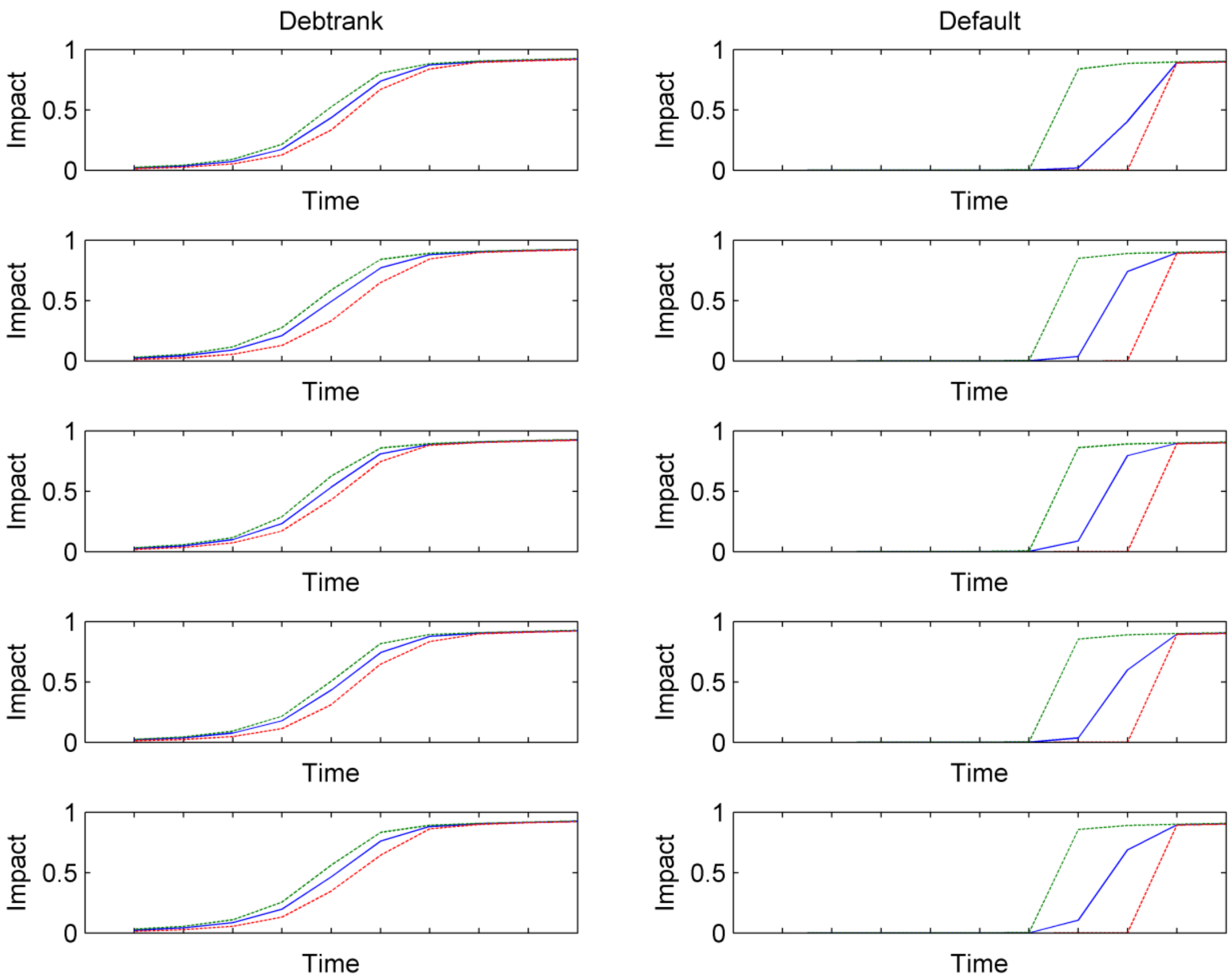

**Figure 4.4 Evolution of Debtrank and default contagion for the five largest banks for the ensemble with $\sum k$ = 3000**

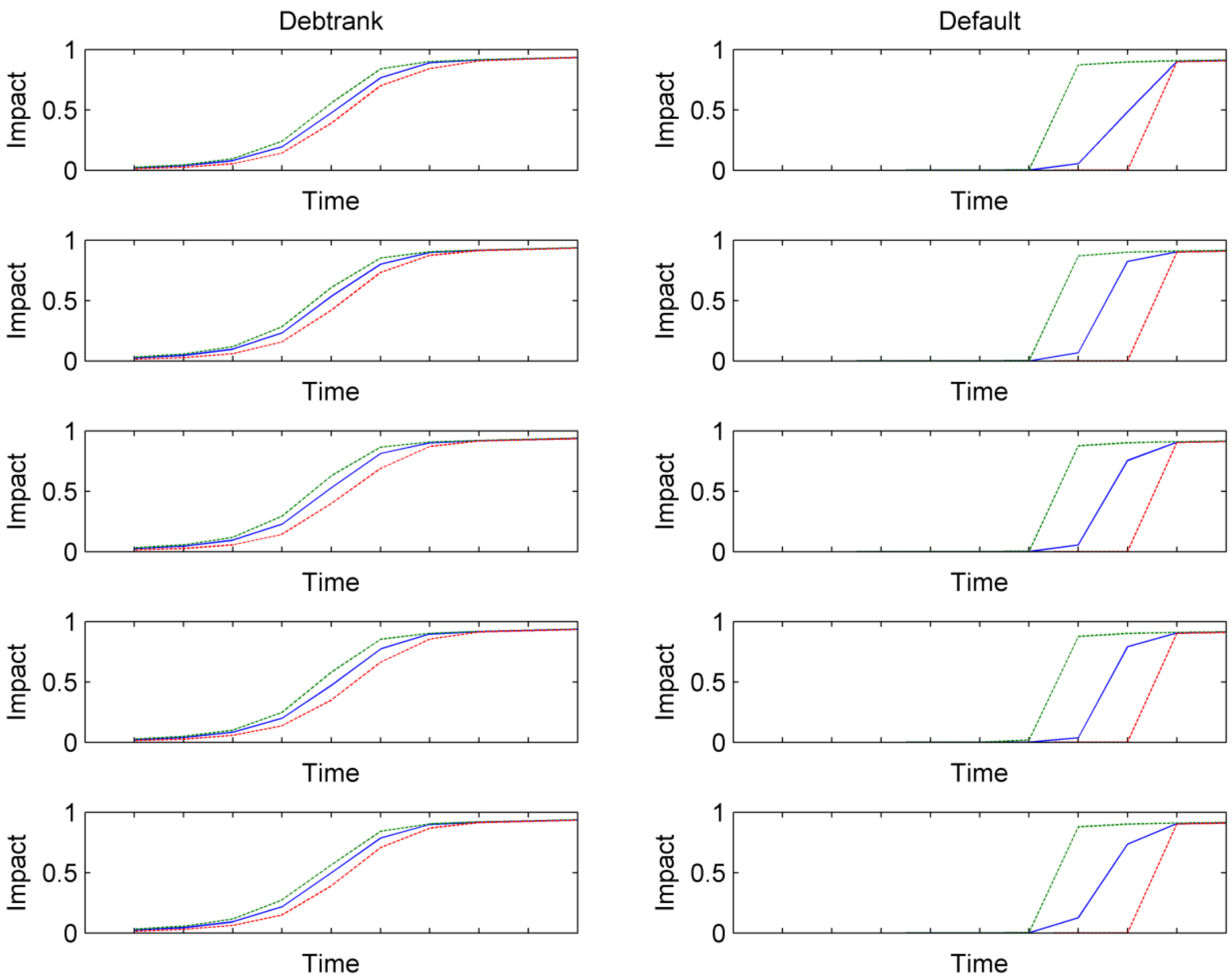

**Figure 4.5 Evolution of Debtrank and default contagion for the five largest banks for the ensemble with $\sum k$ = 5000**

Point (iv) becomes more apparent by analyising how the qualitative properties of the distribution of the impacts estimated by the algorithms changes over time. As a demonstration, figures 4.6 and 4.7 show the variation of this quantity for node 1 in the third ensemble. The default contagion algorithm features zero of extremely minute values until time step 6, after which it starts to have a quasi-binary distribution. Only at the final two time steps it generates a more continuous range of values. The same does not hold for DebtRank, which at every time step exhibits a continuous distribution of variables, making the construction of the uncertainty intervals a more meaningful exercise.

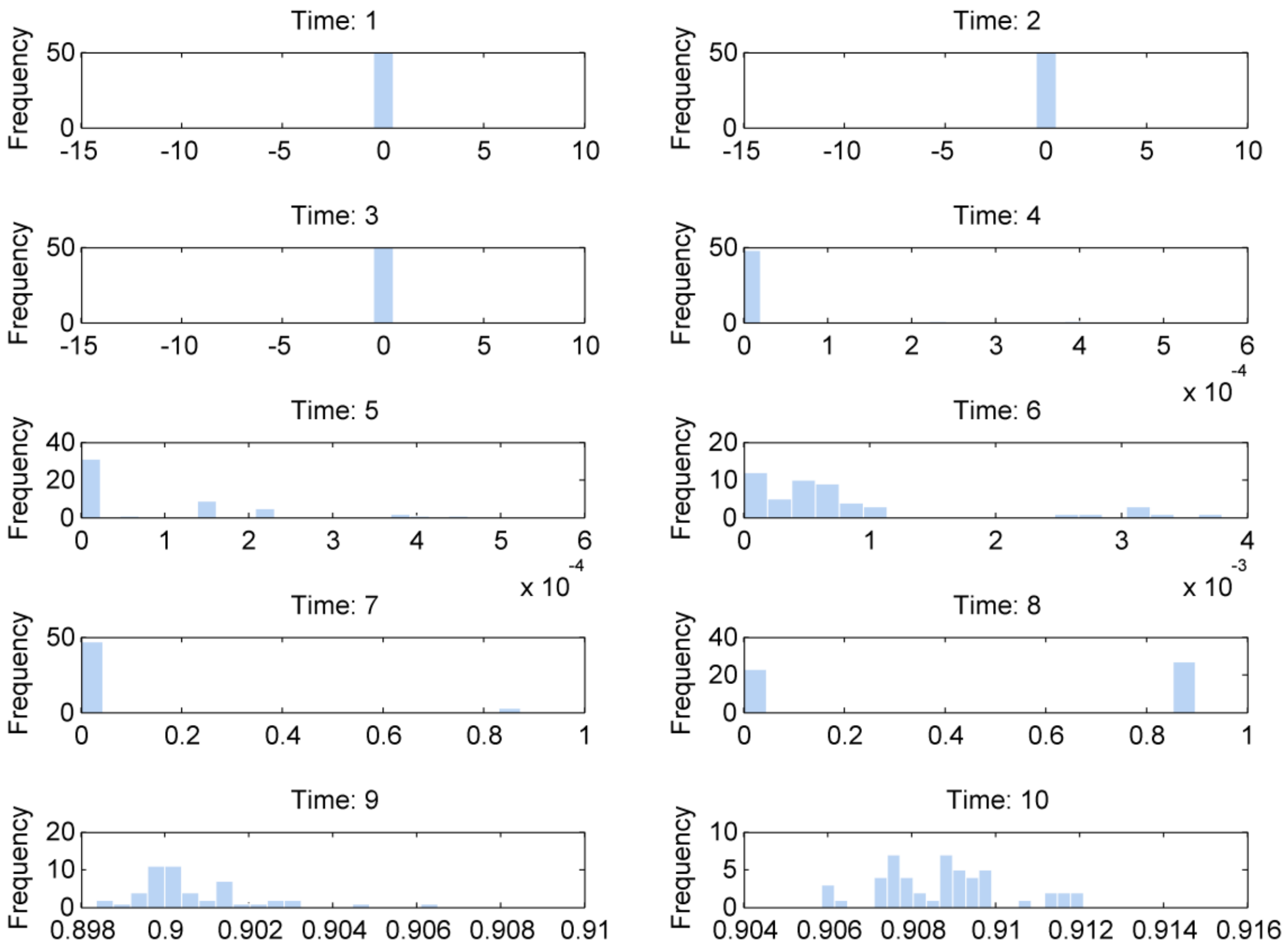

**Figure 4.6 Frequency distribution of the impact evaluated by the default contagion algorithm for node 1 in the third ensemble**

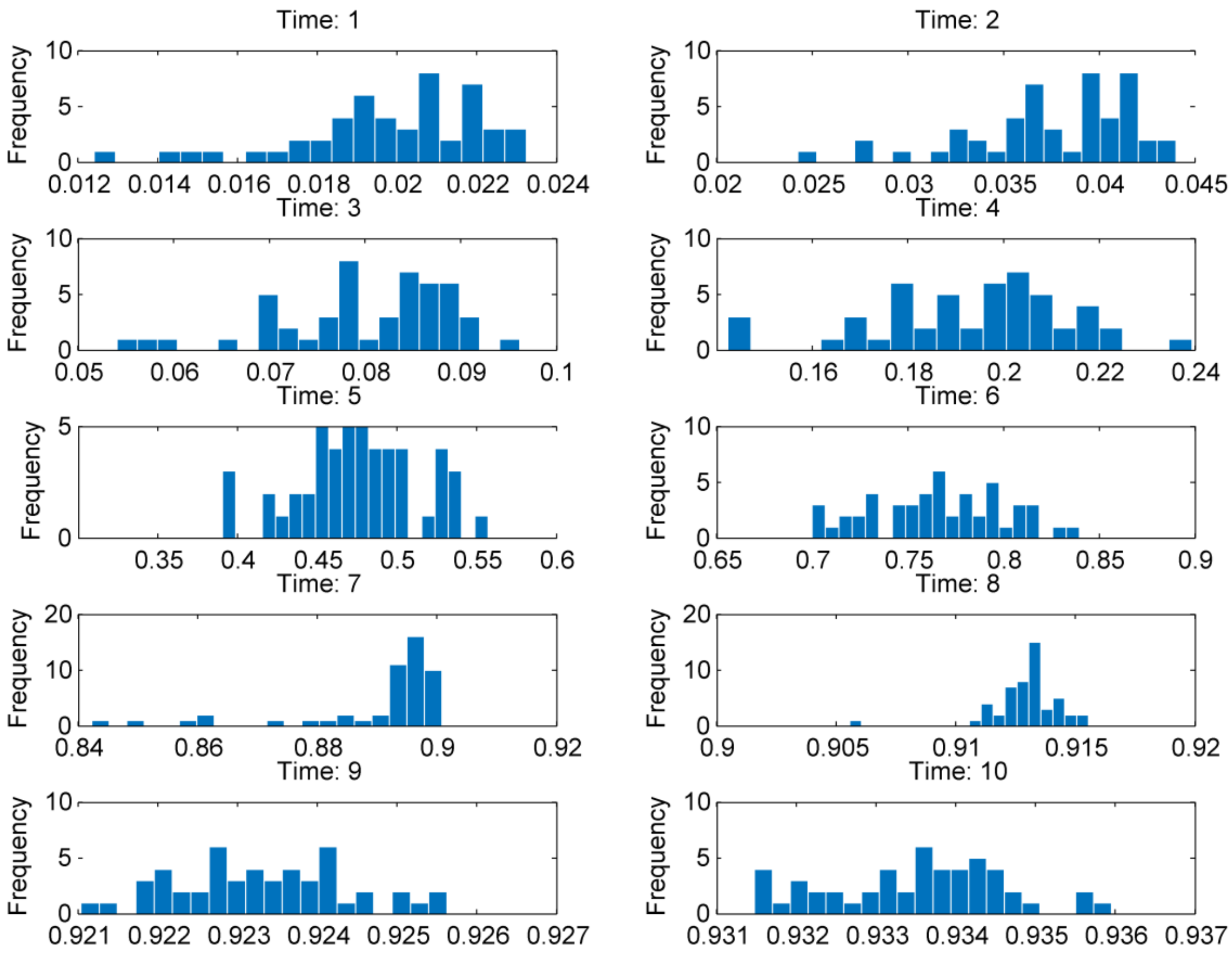

**Figure 4.7 Frequency distribution of the impact evaluated by DebtRank for node 1 in the third ensemble**

## 5. Policy implications

The experiment in Chapter 4 has shown that DebtRank augmented with uncertainty intervals is a suitable candidate for the investigation of how financial systems are affected by adverse events, resulting superior to an alternative financial shock contagion algorithm. The implications of this conclusion are extremely relevant for macroprudential policy making. Indeed, the model can yield a number of instruments that would allow the regulators to assess attentively the conditions of a banking system. Four of such instruments are briefly mentioned in these paragraphs.

The first module alone may already represent an extremely valuable tool. Because of the persistent lack of complete data in this area, a method to reconstruct financial networks can be implemented in many empirical cases. The Bank of England, for instance, features interbank data related to UK-located banks, but not to others headquartered in other countries. This very missing part of the network will be estimated during the empirical validation of the work presented in this work, as shown by the next Chapter. It must also be noted that this part of the model is already in use at the Central Bank of Canada. This only but emphasises the argument made here.

Certainly, the model of this work can also be used to dynamically assess the potential systemic impact of banks over time. In the same fashion of the synthetic experiment, it is possible to reconstruct ensemble of networks and evaluate DebtRank at each quarter, for instance. This would allow the identification of early warning signals, thereby suggesting policy makers what institutions would need closer attention or immediate preemptive action.

A third policy instrument would involve the stress-testing of a financial institution under different shock conditions. Because DebtRank does not necessarily require shocks engendering the complete failure of a bank, it is possible to investigate other conditions of distress. An example is provided in fig. 5.1 below. In this exercise, node 1 of the ensemble with an average of 3000 edges is stress-tested under 20 scenarios at time step 9. At each scenario the shock to be withstood by the bank increases gradually from 0.1 (implying a loss in the value of total assets by 10%) to 1 (total loss of value and, therefore, default). Green and red lines are respectively upper and lower bounds of the impact, while the blue line is the average. This kind of exercise can be replicated by using different distribution of shock value, and on one or more banks altogether.

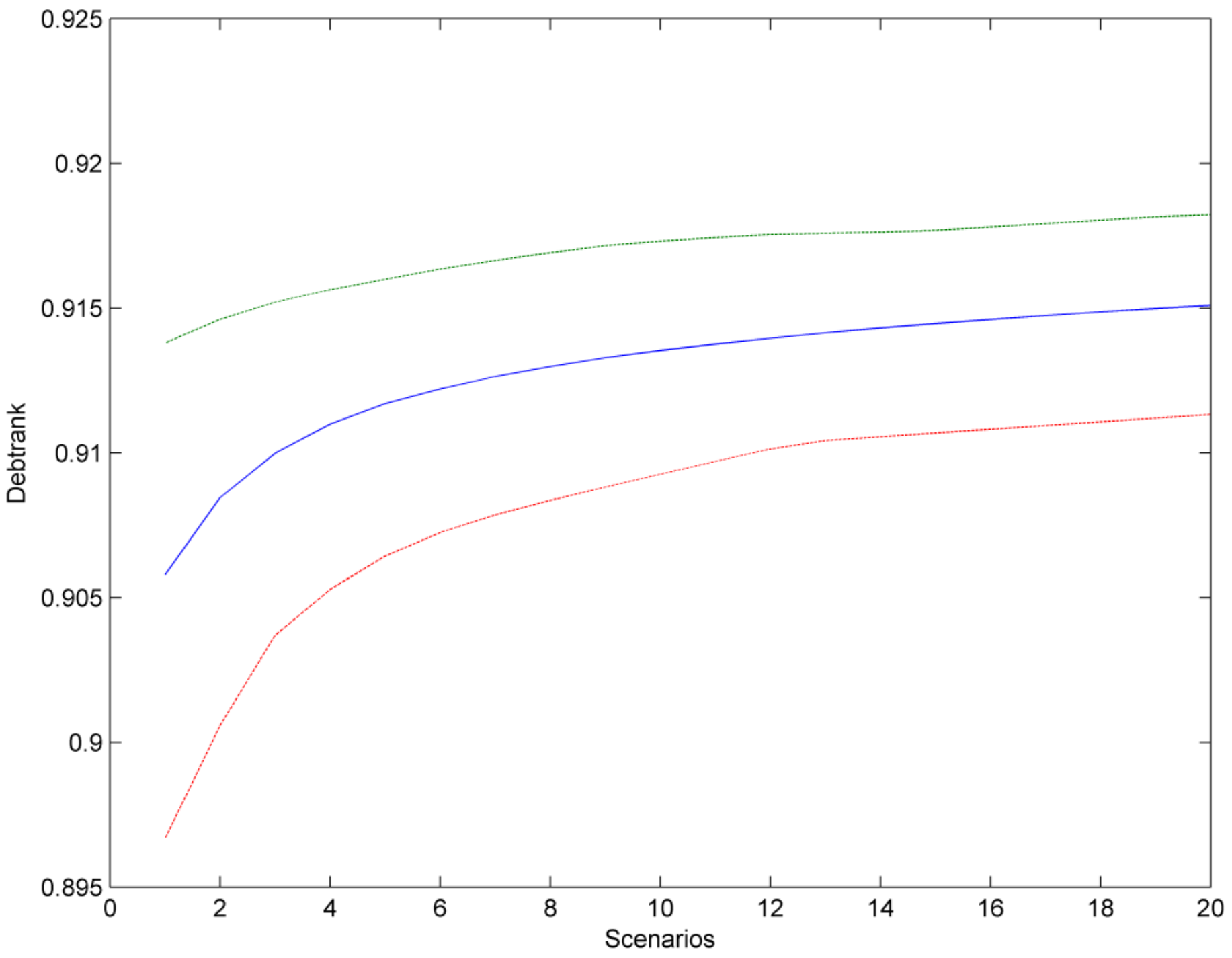

**Figure 5.1 Stress-testing of node 1 at different shock scenarios**

A further and, possibly, more interesting policy application involves the quantification of the losses due to the distress of one or more banks. These can be easily evaluated by quantifying the consequences of the shock in terms of market capitalisation lost[9]. Say, for instance, that it is intended to know how much capitalisation node 1 would lose if node 10 was hit by a pre-determined distribution of shocks, such as the one presented in fig. 5.2. Once generate a matrix with random shocks extracted by the distribution (in this case, it is a truncated normal distribution with mean 0.5 and standard deviation 0.05), it suffices to normalise the impact on node 1 by its market capitalisation value. The result is in fig. 5.3. The losses can be evaluated also for a group of nodes or for the entire network, as demonstrated by figures 5.4 and 5.5. Additionally, the distribution of losses can be evaluated conditional to a group of banks being in distress, not just one.

The elasticity of this method allows the in-depth investigation of individual fragilities of a bank or of a group, by explicitly taking into account their interconnections with the entire system. The method, furthermore, enables the evaluation of Value at Risk (VaR) estimates, a tool that is already widely used by both regulators and the private sector. VaR measures the maximum potential loss of a financial value over a pre-determined time period within a confidence interval (Linsmeier and Pearson, 2000; Jorion, 2006). Analytically, it is defined as below (Artzer et al., 1999):

$$VaR_{\alpha}(L) = inf\{l \in \mathbb{R} : (\mathbf{P} > \mathbf{l}) \leq \mathbf{1} - \alpha\}$$

**Equation 5.1 Value at Risk**

[9] Other ways of quantifying the losses, though, may well be acceptable. Possible instances include levels of total asset values burnt by the financial cascade, or losses in Tier 1 Capital.

Where:

$L$ is the underlying profit and loss probability distribution;
$\alpha$ is a confidence level whose value is within the range [0,1];
$l$ is the value at risk, or the minimum possible value of L such that the likelihood of losses exceeding $l$ is $(1 - \alpha)$.

In the losses distributions below, a VaR with $\alpha = 0.95$ can be easily calculated, and would represent the maximum amount of losses in market capitalisation that would be experienced, with a 95% probability.

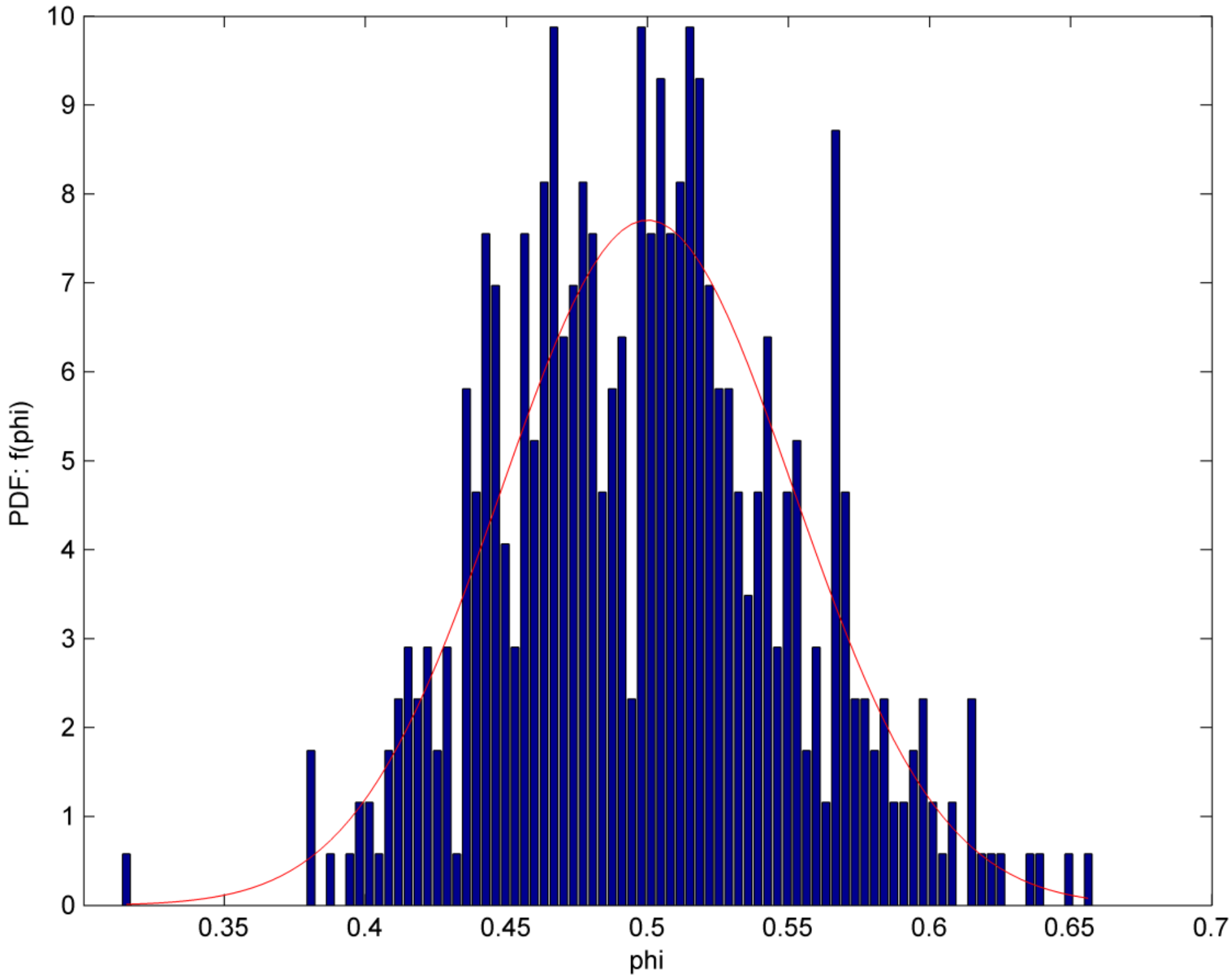

**Figure 5.2 Random shocks on node 10, extracted from $N \sim (0.5, 0.05)$**

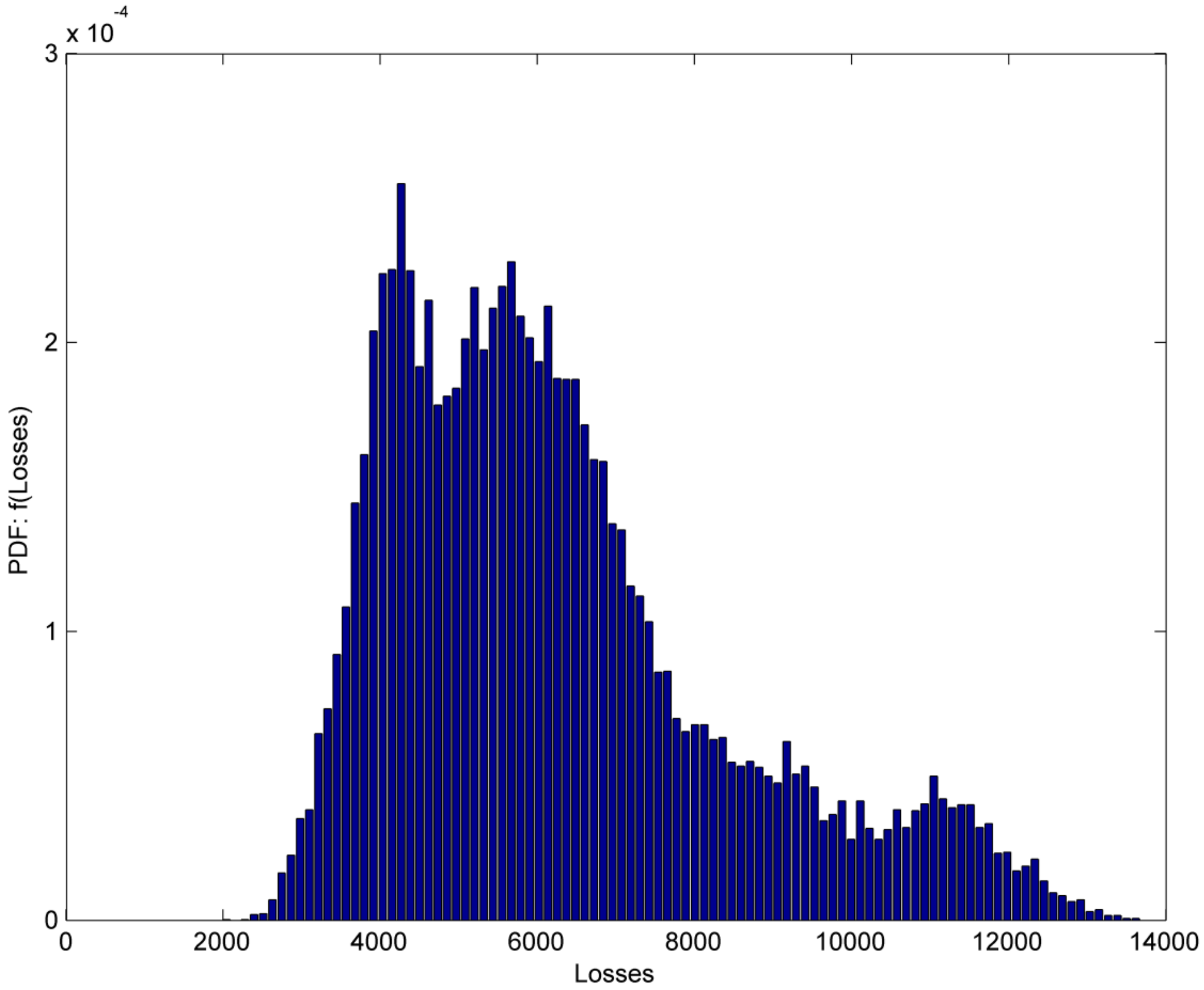

**Figure 5.3 Losses distribution of node 1, due to the shocks simulated on node 10**

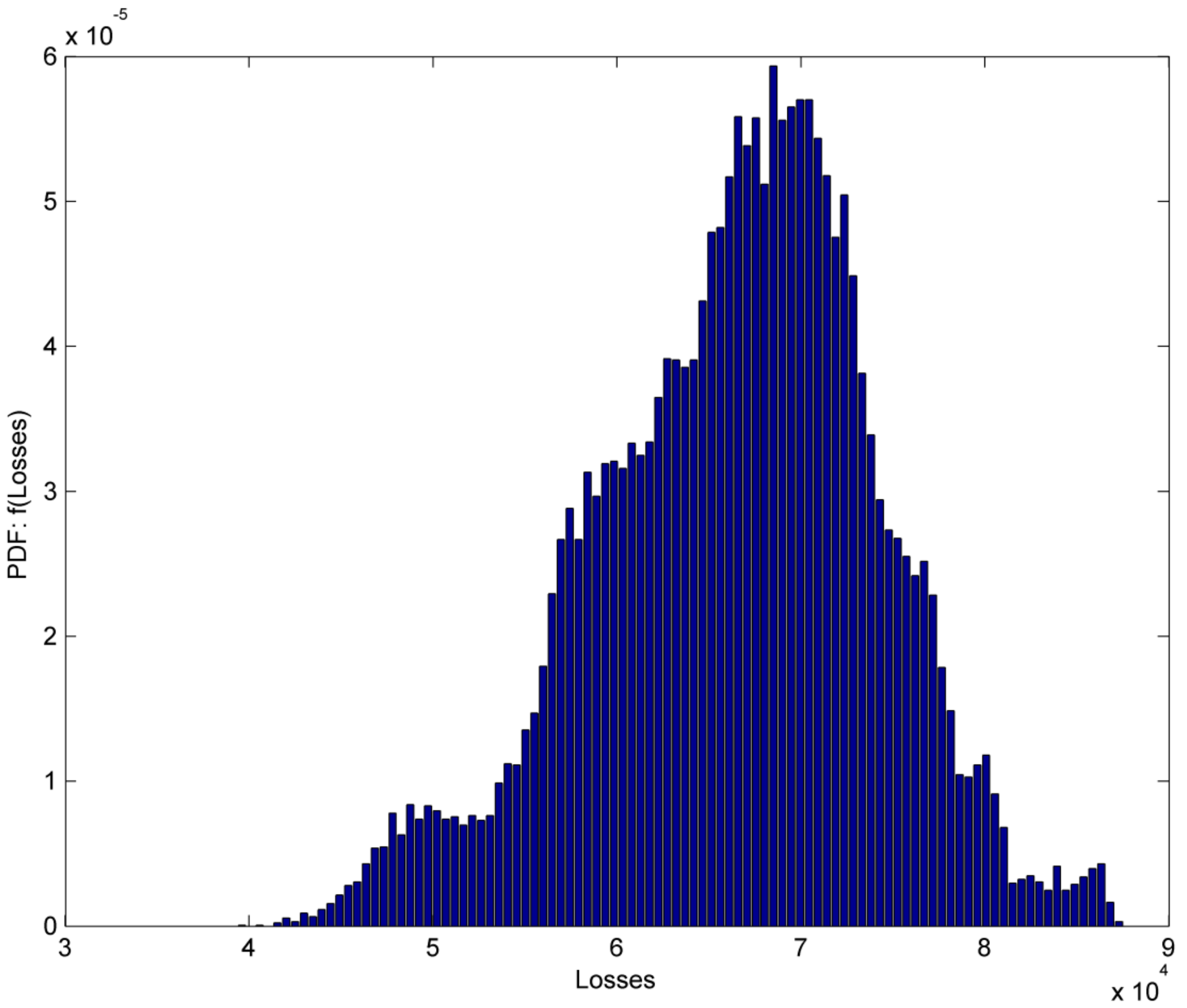

**Figure 5.4 Losses distribution of nodes 1 to 9, due to the shocks simulated on node 10**

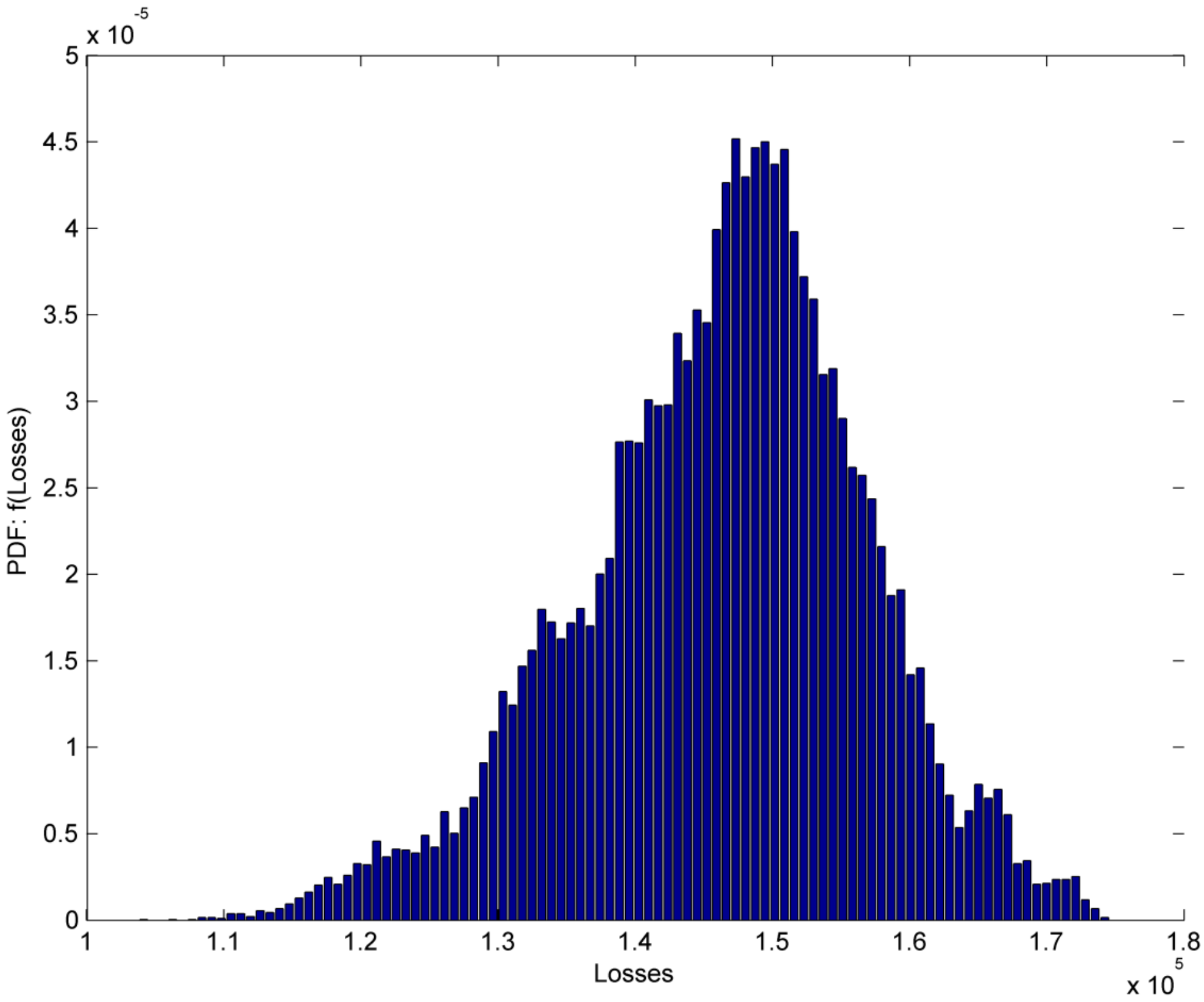

**Figure 5.5 Losses distribution of the entire financial network, due to the shocks simulated on node 10**

Despite the valuable policy implications, the method and results of this paper still suffer from limitations, with a need for further work. These issues are tackled by the next Chapter.

## 6. Limitations and way forward

This work is not immune from limitations. This chapter elaborates three main shortcomings, respectively dealing with the validity of the uncertainty intervals, the lack of validation on a real world setting, and the presence of multiple types of interbank connections.

*Uncertainty intervals.* The construction of ensembles composed of 50 networks has allowed the estimation of useful intervals for the impacts of the shocks simulated in the experiment. Yet, it is most likely that further research ought to feature ensembles with higher number of networks. This is mainly due to the uncertain form which the asymptotic density distributions of the ensembles possess. Larger ensembles would make the form of the distribution clear, thereby suggesting more mathematically rigorous ways of constructing the intervals. This has not been possible in this paper, primarily due to the increasing computational power needed to evaluate the shock impacts as the number of networks in the ensembles increases. Future work will overcome this issue by making use of more computationally powerful machines.

*Validation in a real world setting*. The model is yet to be validated on real-world networks, or, at least, on networks reconstructed not simply from node attributes – as in this paper –, but also from partial network information. This, obviously, sheds some uncertainty as to whether the results gathered in this work apply on fully empirical financial networks. This issue will be addressed by using the model on UK interbank networks, estimated thanks to the information that the Bank of England conserve. This will not only allow the use of these methods for policy purposes, but it will also help adapting the model to real-world contexts. This task is planned to be completed over the second trimester of Year 2 of the PhD program, as exhibited in table 6.1.

*Multiplex networks.* Banks are not connected to each other simply by lending. In financial systems, there have been identified at least three levels of interactions (although, surely, a comprehensive analysis would find more): interbank lending (Iori, Jafarey and Padilla, 2006), overlapping portfolios (Caccioli et al., 2012), and derivatives contracts (Haldane, 2009). Without taking into account all these layers, a truly empirically valid answer to the research question cannot be found. Assuming information about such levels, it is possible to construct a multilayered network of banks, and simulate the propagation of one or more initial adverse events throughout the system, taking into account the different feedbacks that each layer engenders.

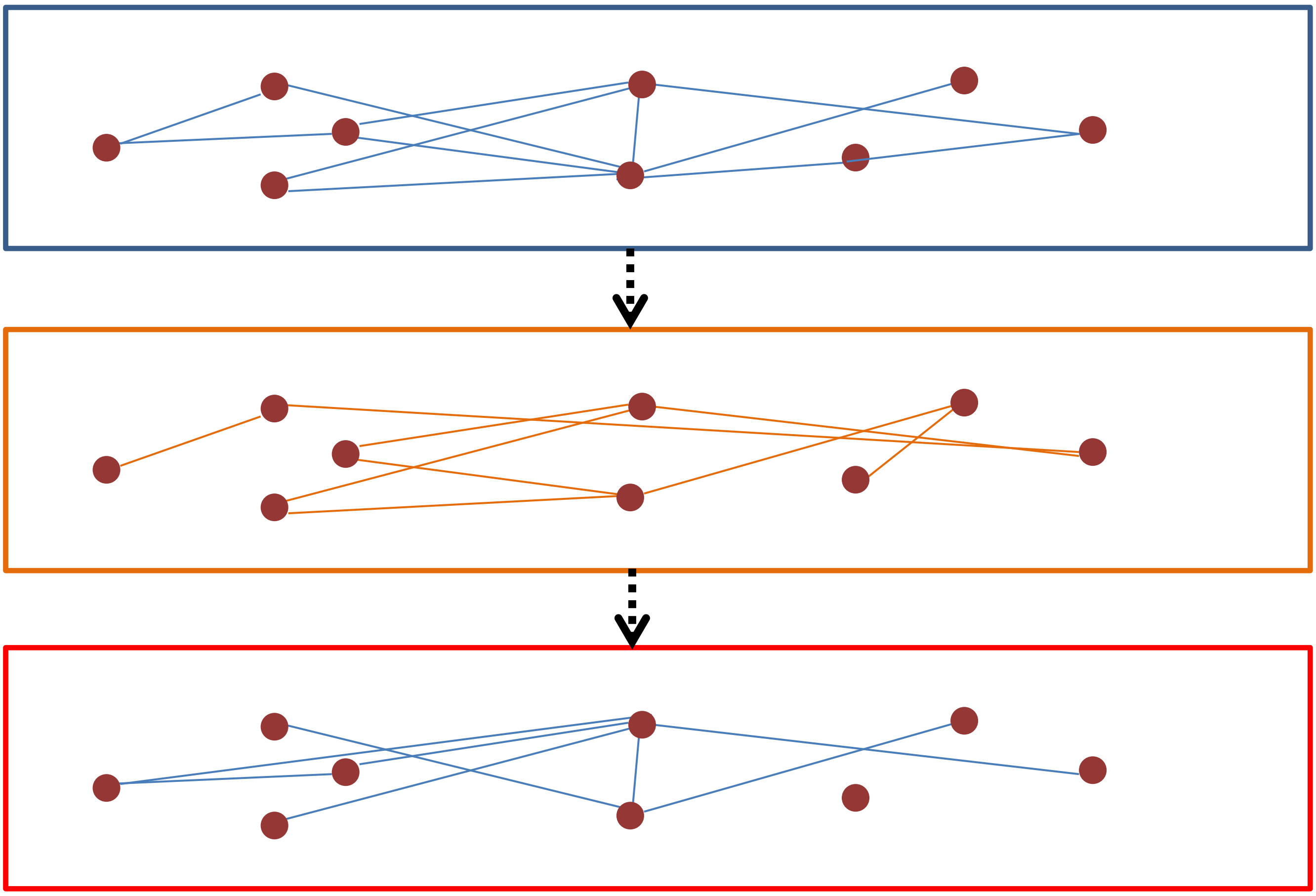

**Figure 6.1 Depiction of a multilayer or multiplex network**

In order to quantify the systemic impact of shocks, a contagion algorithm that takes into account multiplex networks' structures and dynamics should be applied. This method has not yet been developed, but it has been explored theoretically (Yağan and Gligor, 2012). Ideally, the Debtrank algorithm can be augmented to satisfy these conditions. As De Domenico et al. (2013) demonstrate, any mathematical formulations of conventional networks can be transposed to the greater dimensionality of multiplex networks by the use of tensors, geometric objects that allow to synthesise relations between vectors and matrices (Aris, 2012). Debtrank could simply be ameliorated by the insertion of tensor mathematical formulas of contagion. Given the data and the model used, it would be possible to assess how financial systems are affected by shocks with good precision and low levels of uncertainty. The inclusion of all modes of banks' interaction would allow an informed simulation of the propagation of the shock through the system. This task is set to be explored over the last trimester of Year 2 and the first of Year 3.

On another note, it is important to also provide an external validation of the model on novel socio-economic systems, thus shedding more light onto how shocks propagates and impact them. A possible application domain is the world trade web.

## 7. Concluding remarks

This paper has attempted to contribute to the study of financial shocks and contagion, with a major focus on the possibility to quantify how financial systems are affected by extreme adverse events. By adopting a complex networks framework, the research has advanced the use of DebtRank as an algorithm to quantity the impact of the distress of one or more institutions onto the wider system. The algorithm has been augmented with uncertainty intervals, and its performance has been successfully tested against an alternative method, the default contagion algorithm.

The work has arrived to this contribution through the following steps. First, Chapter 2 has introduced the reader to the research question, explicating the consequences of financial crises on GDP growth and unemployment as part of the reasons behind its importance. It has then reviewed three main branches of the relevant literature. Indicator-based measures have been introduced. It has been shown that they make use of (mostly) publicly available data to devise indices stating how important a single financial institution is for the stability of the financial system. Nevertheless, it has been demonstrated that the methodology to construct this type of indices is not able to quantitatively capture the systemic effects of a market shock. Attention has then been shifted towards DSGE models. It has been argued, though, that the very assumption of general equilibrium and the poor complexity of the simulations make this approach fail in providing answers to the research question. As an alternative to the aforementioned works, complex networks analysis has been advanced. It has been stressed that the main strength of this approach is the recognition of the complex and nonlinear nature of financial systems dynamics, which allows for the creation of appropriate models and policy tools, such as the DebtRank algorithm.

Chapter 3 has presented the methods behind the model. After having described the main network concepts used in the framework and having operationalised the main variables of interest with real-world data, the chapter has provided an in-depth account of the main two modules of the model. The first module has the task of reconstructing networks from partial information, by the use of the fitness model. The algorithmic flow of the module has been explicated. The second module applies DebtRank and the default cascade algorithm on the ensembles estimated by the first module, thereby evaluating the impact of simulated shocks. The mathematical framework behind the two algorithms has been shown.

Chapter 4 has been focused on the design of the experiment and on its results. The experiment involved the testing of DebtRank and the default contagion algorithm on three ensembles of 50 networks with different densities, and gradually worse systemic conditions. It has been concluded that DebtRank performs better than the latter algorithm, as it is able to capture early warning signals of the gradual deterioration of the system’s conditions. On the other side, the default contagion algorithm failed to anticipate any distress, and provided impact values which are not adapt for the construction of uncertainty intervals.

Chapter 5 has discussed the policy implications of the findings of the paper. It has been suggested that three policy applications can potentially stem from this model, namely the monitoring of the systemic importance of banks over time, the stress-testing of one or more banks under a distribution of shock scenarios, and the quantification of losses endured by one or more financial institutions, conditional to the distress of other institutions. The latter application has been deemed particularly useful for the identification of Value at Risk values for financial systems.

Finally, Chapter 6 has briefly overviewed the main limitations of the study. It has been contended that the construction of uncertainty intervals would benefit from larger ensembles, and that the model needs validation on real-world financial networks, such as the UK interbank market. Furthermore, it has been recognised the need to investigate financial contagion in the realm of multiplex networks, because financial institutions, after all, interact and affect each other on several different layers. The Chapter has also outlined a potential time schedule for the future work to be done by the completion of the PhD program.